\documentclass[12pt, draftclsnofoot, onecolumn]{IEEEtran}

\usepackage{graphicx,epsfig,amssymb,amsmath,url,latexsym,stfloats,array,bm,bbm}
\usepackage[tight,footnotesize]{subfigure}
\usepackage{cite}
\usepackage{makecell}
\usepackage{mathrsfs}
\usepackage{amsthm} 
\usepackage{longtable}
\usepackage{amsfonts}
\usepackage{caption}
\usepackage{enumitem}
\usepackage{graphicx}
\usepackage{multirow}
\usepackage{longtable}
\usepackage{float}
\usepackage{afterpage}
\usepackage{subeqnarray}
\usepackage{epstopdf}
\usepackage{empheq}
\usepackage{latexsym}
\usepackage{CJK}
\usepackage{color, soul}
\usepackage{framed}
\usepackage{lipsum}
\usepackage{color}
\usepackage{xcolor}
\usepackage{xpatch}
\usepackage{stfloats}
\usepackage{setspace}
\usepackage{dsfont}
\definecolor{shadecolor}{rgb}{1,0,0}

\usepackage{algorithmic}
\usepackage[linesnumbered,ruled]{algorithm2e}

\begin{document}
\title{Digital Twin-Empowered Network Planning for Multi-Tier Computing}
\author{ Conghao~Zhou,~Jie~Gao,~Mushu~Li,~Xuemin~(Sherman)~Shen,~Weihua Zhuang

\thanks{C.~Zhou, X.~Shen, and W.~Zhuang are with the Department of Electrical and Computer Engineering, University of Waterloo, Waterloo, ON, N2L~3G1, Canada (e-mail: c89zhou@uwaterloo.ca; sshen@uwaterloo.ca; wzhuang@uwaterloo.ca).}
\thanks{J.~Gao is with the School of Information Technology, Carleton University, Ottawa, ON, K1S~5B6, Canada (email:~jie.gao6@carleton.ca ).}
\thanks{M.~Li is with the Department of Electrical, Computer, and Biomedical Engineering, Toronto Metropolitan University, Toronto, ON, M5B~2K3, Canada (e-mail:~mushu1.li@ryerson.ca)}

}	
\maketitle

\begin{abstract} 

In this paper, we design a resource management scheme to support stateful applications, which will be prevalent in 6G networks. Different from stateless applications, stateful applications require context data while executing computing tasks from user terminals (UTs). Using a multi-tier computing paradigm with servers deployed at the core network, gateways, and base stations to support stateful applications, we aim to optimize long-term resource reservation by jointly minimizing the usage of computing, storage, and communication resources and the cost from reconfiguring resource reservation. The coupling among different resources and the impact of UT mobility create challenges in resource management. To address the challenges, we develop digital twin (DT) empowered network planning with two elements, i.e., multi-resource reservation and resource reservation reconfiguration. First, DTs are designed for collecting UT status data, based on which UTs are grouped according to their mobility patterns. Second, an algorithm is proposed to customize resource reservation for different groups to satisfy their different resource demands. Last, a Meta-learning-based approach is developed to reconfigure resource reservation for balancing the network resource usage and the reconfiguration cost. Simulation results demonstrate that the proposed DT-empowered network planning outperforms benchmark frameworks by using less resources and incurring lower reconfiguration costs.

\end{abstract}


\newpage
\section{Introduction}

The sixth generation (6G) networks are expected to support a wide range of computing applications~\cite{OJVT_Shen_2020}. A large portion of these applications are \emph{stateful}, meaning that context data is required to execute computing tasks~\cite{shillaker2020faasm,barcelona2019faas}. For example, augmented reality applications require volumetric media objects or holograms, as the context data, to process video segments for user terminals (UTs). The prevalent mobile edge computing (MEC) paradigm provides a solution to supporting computing applications with low offloading delay but has limitations in supporting stateful applications~\cite{kekki2018mec}. Specifically, edge servers close to UTs generally have limited storage capacity to store all context data of stateful applications. Moreover, even if the context data could be fully stored at an edge server, limited communication coverage and a relatively small number of UTs served by the edge server would degrade storage resource utilization.

To address the above limitations, both the industry and the academia have started looking into the collaboration of servers~\cite{krol2019compute, IETF_CFN}. Extending from MEC, \emph{multi-tier computing} integrates multiple servers deployed at the core network, gateways, access points, and other locations in the network for executing computing tasks from UTs. Servers at different tiers have diverse features in terms of resource capacity and service coverage~\cite{wang2022joint}. Specifically, servers deployed at the core network and gateways have larger service coverage and more abundant resources than servers deployed at access points. Through coordinating servers at different tiers, multi-tier computing can exploit different features of servers to support computing applications, especially the stateful ones, in the era of 6G. 

Network planning, as an important part of network management, can facilitate the coordination of servers at different tiers. Network planning consists of resource reservation and resource reservation reconfiguration~\cite{OJVT_Shen_2020}. Resource reservation refers to proactively reserving network resources for satisfying the upcoming resource demands from UTs. Resource reservation reconfiguration refers to timely updating resource reservation decisions to adapt to time-varying resource demands and dynamic network environments. Network planning for supporting stateful applications faces four challenges. First, the reservation of computing, storage, and communication resources for stateful applications is tightly coupled, yielding existing resource reservation solutions for supporting stateless applications inapplicable. Second, the requests for context data may vary across a network, rendering both computing task assignment and storage resource reservation dependent on specific servers and UT mobility patterns. Third, information regarding individual UT status, e.g., UT mobility, is unavailable at the time of network planning, yet such UT-level information can be useful for accurately calculating the amount of network resources needed for supporting stateful applications~\cite{wu2021learning}. Fourth, reconfiguring computing and storage resource reservation for stateful applications in a dynamic network environment yields additional costs due to computing service interruption, which complicates resource reservation reconfiguration~\cite{hwang2014scale}. Addressing the above challenges is important to accurate and adaptive network planning for supporting stateful applications in 6G.

Recently, the digital twin paradigm has started attracting attention as a potential solution to advancing network management for 6G~\cite{shen2021holistic}. The concept of digital twins (DTs) originates from product life-cycle management in industry, where a DT is a synchronized virtual replica of a physical object~\cite{Summit_Yu_2020, TII_Fei_2018}. For 6G networks, DTs can be introduced to represent individual UTs. Each DT consists of a \emph{UT data profile} that describes the corresponding UT, including the UT's mobility, service demands, and quality of service (QoS) satisfaction, and \emph{DT functions} for data acquisition, processing, and analysis~\cite{shen2021holistic}. The introduction of DTs brings three benefits to network planning. First, historical data contained in DTs can be used to predict UT status in the upcoming time interval, which can, in turn, facilitate customized resource reservation for highly diversified UTs and enable fine-grained network planning. Second, data indicating the performance of network planning can be collected based on DTs, which can provide a foundation for resource reservation reconfiguration in network planning to adapt to a highly dynamic network environment. Third, DTs should acquire extensive and well-organized data that can be used to explore and exploit hidden network characteristics, thereby facilitating effective data-driven network planning approaches to enhancing network performance. Due to the above benefits, DTs can be exploited and designed to improve the granularity, adaptivity, and intelligence of network planning in 6G.

In this paper, we design a network planning scheme for supporting a stateful application in the scenario of multi-tier computing. Our research objective is to find out the minimum amount of network resources (including computing, storage, and communication) needed for supporting the application and also balance the resource usage and the cost from resource reservation reconfiguration in a dynamic network environment. To achieve this objective, we propose a DT-empowered network planning framework with the following two elements: group-based multi-resource reservation and closed-loop resource reservation reconfiguration. First, we design DTs for individual UTs to characterize their status and group them based on their mobility patterns. We propose an algorithm based on matching theory and particle swarm optimization to address the coupling relation among computing, storage, and communication in resource reservation. The proposed method enables customized resource reservation for satisfying different resource demands of UT groups with different mobility patterns. Second, we develop a Meta-learning-based approach for resource reservation reconfiguration to cope with the dynamic network environment. The main contributions of this paper are the followings: 
\begin{itemize}
\item We propose a novel network planning framework to facilitate fine-grained resource reservation based on UT data contained in DTs;

\item We address a challenging multi-resource reservation problem for supporting stateful applications in multi-tier computing;

\item We develop an automated closed-loop approach to reconfigure resource reservation in a dynamic network environment for balancing the network resource usage and the cost from reconfiguring resource reservation.
\end{itemize}

The remainder of this paper is organized as follows.
Section~II provides an overview of related studies.
Section~III describes the considered network scenario, the proposed DT-empowered framework, and the system model.
Section~IV formulates the network planning problem for multi-tier computing.
Sections~V and~VI introduce the proposed solutions for resource reservation and resource reservation reconfiguration, respectively.
Section~VII presents the simulation results, followed by the conclusion and future work in Section~VIII.

\section{Related Work}

Network resource management is conducted in both short-term and long-term~\cite{OJVT_Shen_2020}. Short-term resource allocation in the \emph{operation stage} relies on real-time information on individual UTs, such as UT locations, and targets real-time UT satisfaction. By contrast, long-term resource reservation in the \emph{planning stage} is based on aggregated information of UTs, such as the number of UTs covered by an access point, and focuses on network performance, such as resource utilization over a relatively long time period, ranging from several minutes to hours~\cite{wang2016rethinking, VTM_Li_2021}. 

Most works on real-time resource allocation focus on computing task offloading and service placement, among which many consider one-tier computing such as cloud computing or MEC~\cite{chen2014decentralized,he2020edge,pasteris2019service,ouyang2018follow,kuang2019partial}. Based on the real-time computing task arrival of each UT, decentralized and centralized communication and computing resource allocation approaches are proposed to minimize the delay of computing task offloading and executing in cloud computing and MEC, respectively~\cite{chen2014decentralized,he2020edge}. Service placement is studied in MEC based on the real-time UT location and the type of service required by each UT to maximize the number of UTs that can be severed under each edge server's resource capacity~\cite{pasteris2019service,ouyang2018follow}. Some works focus on resource allocation for multi-tier computing~\cite{lera2018availability,chen2019budget,yu2018computation,zhou2020deep,cheng2019space}. Computing resources on fog nodes and the cloud server are allocated to UTs at different locations to satisfy the delay requirements of their computing tasks~\cite{lera2018availability}. Li~\textit{et al.} investigate a service placement approach for cloud and edge computing to satisfy each UT's computing demands~\cite{chen2019budget}. Given that the same type of computing tasks can share computing results, Yu~\textit{et al.} study joint computing task offloading and service placement in multi-tier computing to reduce the delay of executing computing tasks~\cite{yu2018computation}. Considering space-ground integrated networks, the authors in~\cite{zhou2020deep,cheng2019space} investigate how to allocate computing and communication resources available at satellites, terrestrial access points, and UTs to minimize the delay of executing computing tasks from UTs.

There are less studies on long-term resource reservation for computing task offloading. Based on the aggregated computing demands from all access points, a proactive computing resource reservation approach in MEC is designed to minimize the delay of executing computing tasks~\cite{hu2022cec} and maximizing resource utilization in computing task execution~\cite{zhang2021dynamic}, respectively. Yin~\textit{et al.} study edge server placement to minimize the network resource usage based on statistical computing demands~\cite{yin2016edge}. There are limited works on long-term resource reservation in multi-tier computing~\cite{zhou2020iot,wu2021learning}. Considering edge and cloud computing, Zhou~\textit{et al.} propose a computing resource reservation approach to minimize network resource usage while satisfying different delay requirements of two applications~\cite{zhou2020iot}. For servers located at different tiers of space-ground integrated networks, joint communication and computing resource reservation is studied to minimize the long-term cost of delay requirement violation and network reconfiguration~\cite{wu2021learning}.

While allocating resources to UTs to satisfy their real-time computing demands is important, proactively reserving resources on servers is also essential. In this work, we focus on network planning in the presence of the coupling relation among computing, storage, and communication resources and the impact of UT mobility in long-term resource reservation to support stateful applications. We leverage DTs to improve the granularity and effectiveness of network planning as compared to conventional resource reservation.

\section{Network Scenario and System Model}

In this section, we first introduce the considered network scenario of multi-tier computing. Then, we propose a DT-empowered network planning framework to support stateful applications and present the corresponding system model.

\subsection{Network Scenario}

	\begin{figure}[t]
		\centering
	  	\includegraphics[width=0.7\textwidth]{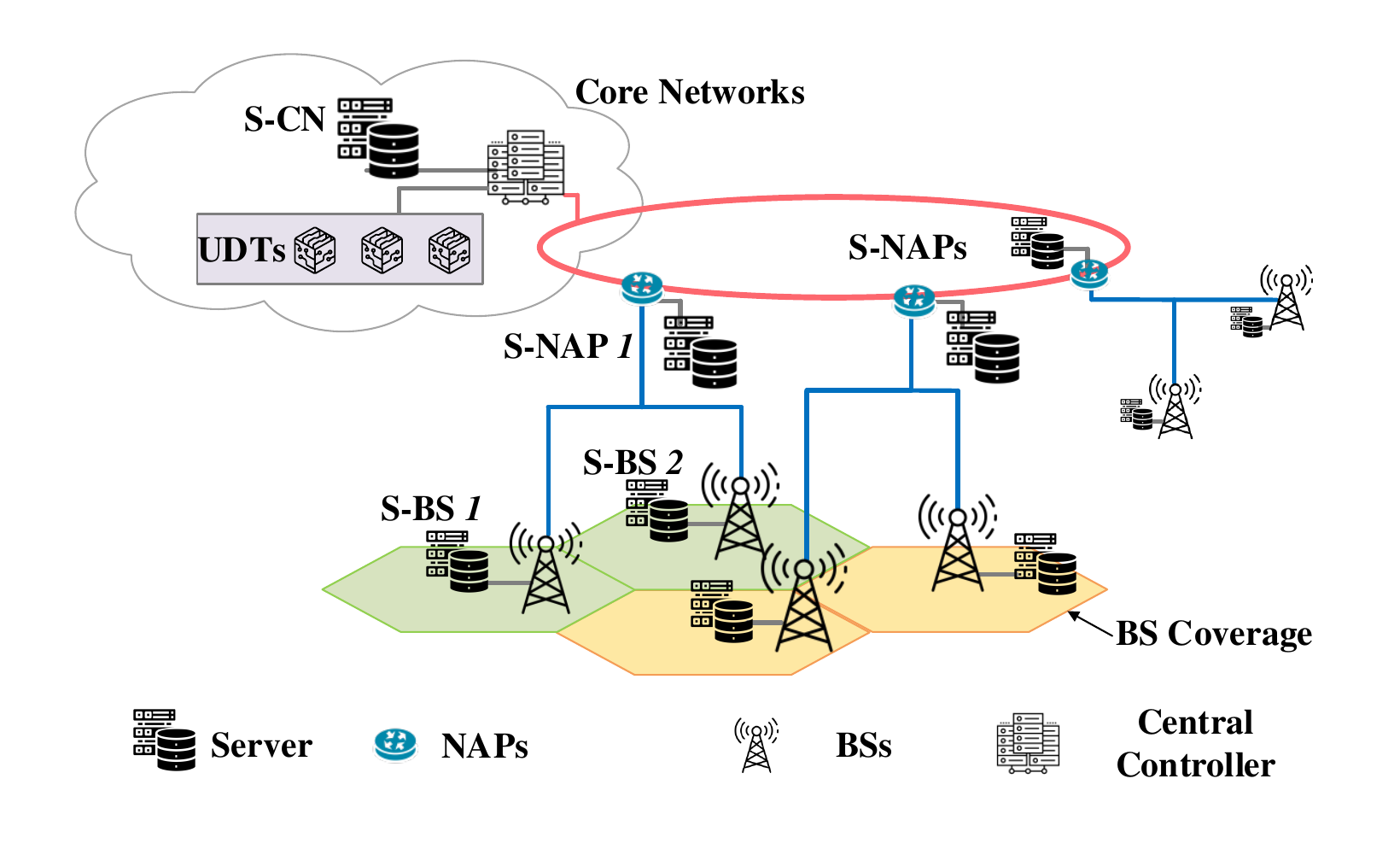}
	  	\caption{The considered scenario of multi-tier computing.}\label{model}
	\end{figure}

The considered scenario is shown in Fig.~\ref{model}. Computing servers are deployed at different locations, including: (i) base stations (BSs); (ii) network aggregation points (NAPs), such as gateways; and (iii) the core network (CN)~\cite{kekki2018mec}. Each server can build one dedicated virtual machine (VM) to execute computing tasks from UTs for the stateful application. The service coverage area of a server at a BS (S-BS) is the BS's communication coverage. The service coverage area of a server at a NAP (S-NAP) is the union of the communication coverage of all the BSs connected to it. For example, S-NAP~$1$ connects to S-BS~$1$ and S-BS~$2$, and the communication coverage area of S-NAP~$1$ consists of the two green cells as shown in Fig.~\ref{model}. The server at the CN (S-CN) can provide computing service to all UTs in the considered network. The service coverage areas of servers at the same tier of the network do not overlap. UTs generate computing tasks and offload their computing tasks to a server when located in its service coverage area. We assume that the computing task generation at all UTs corresponds to an identical statistical process due to the same stateful application. The VM at the server then executes the offloaded computing tasks and sends computing results back to the UTs. Denote the S-CN by~$e^\text{cn}$, and let $\mathcal E^\text{bs}$ and $\mathcal E^\text{nap}$ denote the set of S-BSs and the set of S-NAPs, respectively. Let $\mathcal{E} = \mathcal{E}^\text{bs}\cup \mathcal{E}^\text{nap}\cup \{e^\text{cn}\}$ represent the set of all servers in the network. Denote the set of BSs by $\mathcal B = \{1,2, \cdots, B\}$, and let $\mathcal{E}_{b} \subset \mathcal{E}$ denote the set of servers that include BS~$b \in \mathcal{B}$ in their service coverage areas.

We focus on resource reservation to support stateful applications, which requires context data for executing computing tasks. The application uses a fixed set of context data, and the popularity of different data chunks in this set is different. Moreover, the popularity of each data chunk can vary at different BSs across the network, i.e., the popularity of context data chunks is location-dependent. Therefore, if a UT moves into the coverage of a different BS, its requests for context data may also change, creating a dependence of its resource demand on its mobility. A UT that connects to more BSs within a time interval is considered to have higher mobility. UTs with similar mobility patterns can be grouped together in resource reservation due to their similar need for context data.

The time duration of interest is partitioned into $K$ time intervals of length~$\tau$ (e.g.,~5 to 10 minutes per time interval). Denote the set of time intervals by~$\mathcal K = \{1,2, \cdots, K \}$. A central controller in the CN maintains the DTs for UTs and makes resource reservation decisions proactively for the stateful application at the beginning of each time interval. We define vector $\boldsymbol{\phi}_{d,k}$ to represent the mobility pattern of UT $d$ in time interval $k \in \mathcal{K}$, where the $b$\,th element of $\boldsymbol{\phi}_{d,k}$ represents the time duration in which UT $d$ is in the coverage of BS $b$ during the time interval. Table~I lists the important symbols for easy reference.
 
\begin{table}[t]
\footnotesize 
\centering
\captionsetup{justification=centering,singlelinecheck=false}
\caption{List of Symbols}\label{table1}
\begin{tabular}{c|l}
\hline\hline
 Symbols & \thead{Definition} \\
\hline\hline
$a_{k}$ & \thead[l]{The indicator on whether the resource reservation in time interval $k$ should be reconfigured} \\
\hline
$c_{e,k}^{n}$ & \thead[l]{The amount of computing resource reserved at server $e$ for group $n$ in time interval $k$}\\
\hline
$f_{b,e,k}^{n}$ & \thead[l]{The load of computing tasks from group $n$ covered by BS $b$ and assigned to server $e$ in time interval $k$ } \\
\hline
$f_{b,e,k}^{n,i}$ & \thead[l]{The load of computing tasks requiring chunk~$i$ from group $n$ covered by BS $b$ and assigned to\\ server $e$ in time interval $k$} \\
\hline
$g_{e,k}$ & \thead[l]{The amount of storage resource reserved at server $e$ in time interval $k$} \\
\hline
$\mathcal{I}_{e,k}$ & \thead[l]{The set of chunks stored at server $e$ in time interval $k$} \\
\hline
$L$ & \thead[l]{The data volume of each chunk of context data} \\
\hline
$L^\text{re}$ & \thead[l]{The data volume of remotely accessing context data for each computing task} \\
\hline
$m_{e,k}^{n}$ & \thead[l]{The load of computing tasks from group $n$ assigned to server $e$ in time interval $k$} \\
\hline
$o_{k}^\text{v}$ & \thead[l]{The cost from reconfiguring resource reservation in time interval $k$} \\
\hline
$p_{b,k}^{i}$ & \thead[l]{The request ratio of chunk~$i$ in the coverage of BS~$b$ in time interval $k$} \\
\hline
$v_{b,e,k}^\text{up}$, $v_{b,e,k}^\text{down}$ & \thead[l]{The communication resource usage of uploading and downloading, respectively,\\ between server $e$ and BS $b$ in time interval $k$} \\
\hline
$v_{n, e,k}^\text{re}$ & \thead[l]{The communication resource usage of remote access for executing computing tasks\\ at server $e$ in time interval $k$} \\
\hline
$w^\text{c}$, $w^\text{s}$, $w^\text{o}$ & \thead[l]{The weights of communication, storage, and communication resource usage, respectively} \\
\hline
$x_{b,k}^{n}$, $\tilde{x}_{b,k}^{n}$ & \thead[l]{The actual and predicted load of computing tasks, respectively, from group $n$ in the coverage of\\ BS $b$ in time interval $k$} \\
\hline
$\tilde{\bm{x}}_{k}^{n}$ & \thead[l]{The spatial task distribution of group $n$ in time interval $k$} \\
\hline
$\alpha$, $\beta$, $\gamma$ & \thead[l]{The input data size, computing workload, size of computing result of each computing task, respectively} \\
\hline
$\Delta_{k}$ & \thead[l]{The network resource usage during time interval $k$} \\
\hline
$\epsilon_{e,k}^\text{c}, \epsilon_{e,k}^\text{s}$ & \thead[l]{The computing and storage resource usage of server $e$, respectively, during time interval $k$} \\
\hline
$\eta_{b,e}$  & \thead[l]{The communication resource usage of transmitting a bit data between BS $b$ and server $e$}\\
\hline
$\xi_{e}^{e'}$  & \thead[l]{The communication resource usage of accessing context data remotely from server $e'$ to server~$e$}\\

\hline\hline

\end{tabular}
\end{table}

\subsection{DT-empowered Network Planning }

\begin{figure}[t]
		\centering
	  	\includegraphics[width=0.55\textwidth]{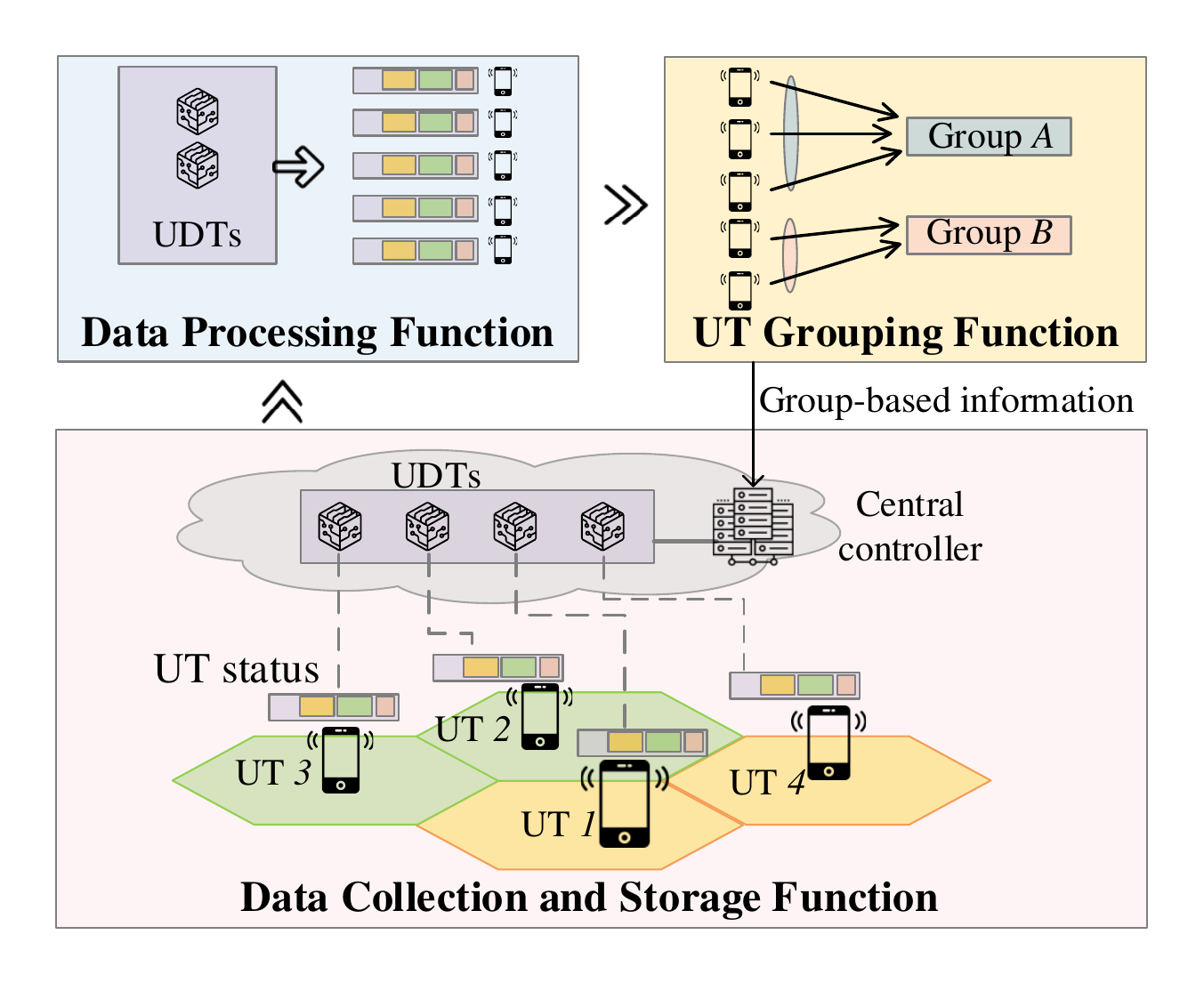}
	  	\caption{The designed UDTs used for network planning. }\label{udt}
	\end{figure}

In this subsection, we present the idea of DT-empowered network planning with our specific design of DTs, which is a succession and development of the framework in~\cite{shen2021holistic}. A DT is created for an individual UT, called UDT, which consists of a UT data profile and UDT functions. As shown in Fig.~\ref{udt}, UDTs are located at the CN, and UT data profiles are maintained and updated by the central controller via UDT functions. Specifically, each UT data profile is a well-organized set of UT data. In this work, the data attributes of each UDT consist of the mobility of the corresponding UT, including the UT's location and velocity, as well as the information of each computing task from the UT, including context data requirement, input and output data size, computing workloads, and resource demands. There are three UDT functions used to manage and analyze UT data profiles for network planning, as follows:
\begin{itemize}
\item \textbf{Data collection and storage function --} UT data required for network planning is collected from individual UTs via BSs or offered by service providers. Data regarding UT mobility, e.g., UT locations, can be uploaded by individual UTs periodically, as specified in 5G. Data regarding services, e.g., computing workloads and required context data, can be obtained from service providers or computing servers~\cite{kekki2018mec}. The collected UT data is stored in the corresponding UDTs;

\item \textbf{Data processing function --} At the beginning of time interval $k$, the mobility patterns of each UT in the past $T$ time intervals, i.e., $\boldsymbol{\phi}_{d,k'}, \forall k' \in [k-T, k-1]$, are obtained based on the UT data in the corresponding UDT. Then, the historical mobility patterns are used to predict the mobility pattern of each UT, i.e., $\boldsymbol{\phi}_{d,k}$, in the subsequent time interval. Other data prediction, e.g., spatial task distributions and requested context data, based on historical data in UDTs is also conducted via this function;

\item \textbf{UT grouping function --} UTs are grouped based on their predicted mobility patterns, i.e., $\{\boldsymbol{\phi}_{d,k}, \forall d\}$, at the beginning of each time interval. UTs from the same group have similar network resource demands for executing computing tasks due to their similar mobility patterns, allowing for an accurate approximation of resource demands for UTs from each group. Based on UT grouping, network resources can be reserved accurately to achieve fine-grained network planning.\footnote{Note that the mechanism of UT grouping, as well as the data attributes used for grouping, can be customized based on the data contained in UDTs in different network scenarios. The number of groups depends on the trade-off between granularity and complexity.}
\end{itemize}
The UDT functions are used to manage and analyze UT data profiles to empower network planning. However, the UDT functions do not make network planning decisions but only provide information for network planning.

\begin{figure*}
	  \centering
	    \subfigure[Workflow.]
	    { \includegraphics[width=0.6\textwidth]{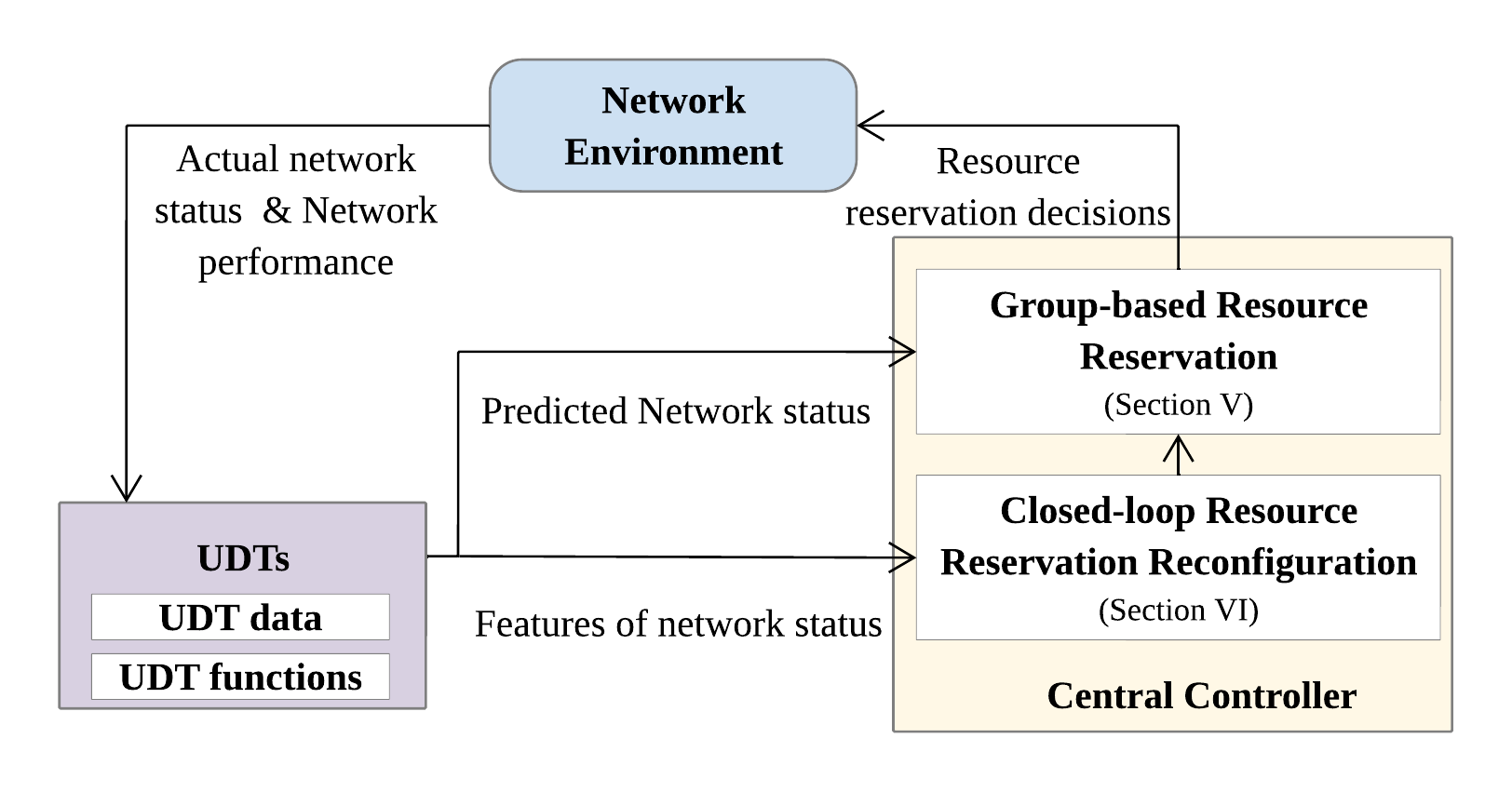}\label{loop} }
		\subfigure[Time line.]
		{ \includegraphics[width=0.6\textwidth]{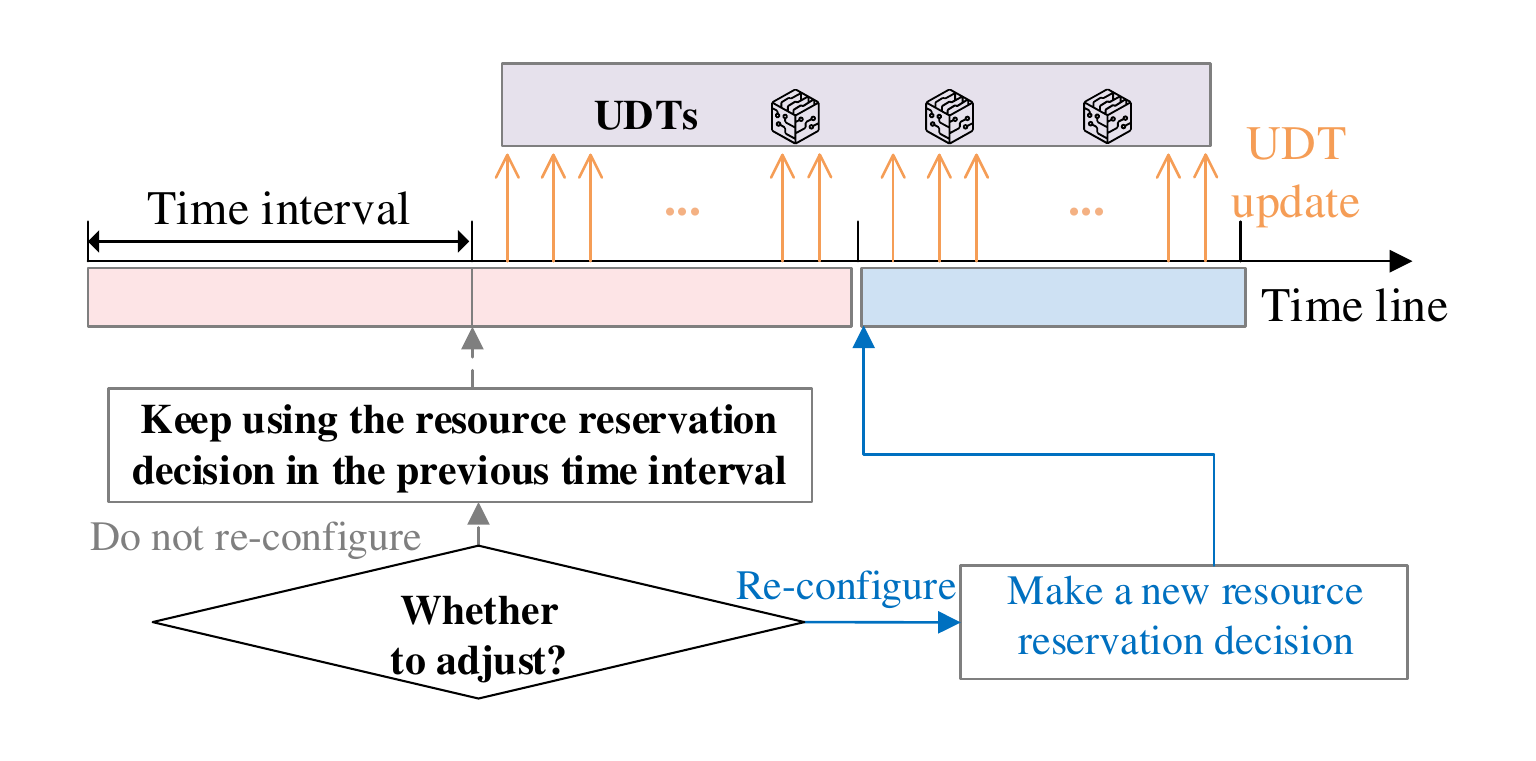}\label{timeline} }
	  \caption{An illustration of DT-empowered network planning.}\label{framework}
	\end{figure*}

Based on UDTs, we propose a novel network planning framework as shown in Fig.~\ref{framework}, which includes two core elements: group-based resource reservation and closed-loop resource reservation reconfiguration.

\begin{itemize}
\item \textbf{Group-based resource reservation:} Based on UT grouping, we reserve storage and computing resources accurately to satisfy the resource demands for UEs in each group in the upcoming time interval. Denote the set of groups by $\mathcal N = \{1,2, \cdots, N\}$. Let~$x_{b, k}^{n}$ denote the number of computing tasks generated by the UTs who are associated with group~$n$ and are in the coverage of BS~$b$ during time interval~$k$. We refer to matrix $\bm{x}_{k}^{n} = [x_{b, k}^{n} ]_{\forall b \in \mathcal{B}}$ as \emph{spatial task distribution} of group~$n$ in time interval~$k$. Since it is impossible to know the actual value of~$\bm{x}_{k}^{n}$ at the beginning of time interval~$k$, $\bm{x}_{k}^{n}$ is predicted based on historical data contained in UDTs. We use superscript~``$\sim$'', e.g.,~$\tilde{x}$, to represent the predicted values of $x$. Given spatial task distributions, we propose an algorithm to address the storage and computing resource reservation problem with highly coupled variables. The detail of group-based resource reservation is discussed in Section~V. 

\item \textbf{Closed-loop resource reservation reconfiguration:} Since spatial task distributions may change across time intervals, reconfiguring the reserved resources on servers to adapt to dynamic computing demands is necessary but yields additional cost from reconfiguring resource reservation. We design a closed-loop approach to reconfigure resource reservation reconfiguration for balancing the network resource usage and the cost from reconfiguring resource reservation, and the time line is shown in Fig.~\ref{timeline}. Specifically, at the beginning of each time interval, the central controller identifies whether to reconfigure resource reservation with a proposed algorithm. If the resource reservation needs to be reconfigured for the subsequent time interval, the controller will make a new resource reservation decision; Otherwise, the controller will keep using the resource reservation from the previous time interval. At the end of each time interval, the data regarding network performance is collected based on UDTs. The detail of closed-loop resource reservation reconfiguration is discussed in Section~VI.

\end{itemize}

\subsection{Computing Model}

The overall load of computing tasks assigned to each server during a time interval affects network resource reservation on the server. Denote the load of computing tasks generated by UTs from group~$n$ in the coverage of BS~$b$ and assigned to server $e$ during time interval~$k$ by~$f_{b,e,k}^{n}$. The relation between computing task assignment and spatial task distribution is given by
	\begin{equation}\label{eq1}	
		\sum_{e \in \mathcal{E}_{b}}{ f_{b,e,k}^{n} } = \tilde{x}_{b, k}^{n}, \,\, \forall b \in \mathcal{B},  n \in \mathcal{N},  k \in \mathcal{K}.
	\end{equation}
Let $m_{e,k}^{n}$ denote the overall load of computing tasks generated by UTs from group~$n$ and assigned to server~$e$ during time interval~$k$,
	\begin{equation}\label{eq2}
		m_{e,k}^{n} = \sum_{b \in \mathcal B_{e}}{f_{b,e,k}^{n}}, \,\, \forall e \in \mathcal{E},  n \in \mathcal{N},  k \in \mathcal{K},
	\end{equation}
where $\mathcal B_{e}$ denotes the set of BSs within the service coverage of server~$e$. In Eq.~\eqref{eq2}, the fine granularity of computing task assignment, reflected through $f_{b,e,k}^{n}$, helps determine the computing load at each server accurately.

Since we consider one specific stateful application, the computing tasks for this application are assumed to have (approximately) the same input data size (in bits), computing workload (in CPU cycles per bit), and result data size (in bits).\footnote{The proposed framework can be straightforwardly extended to handle the case of computing tasks with different input data sizes, computing workloads, and result data sizes by leveraging DTs to collect such information for each computing task. The problem-solving approach (including the proposed algorithms) remains applicable in such a case.} Let $\alpha$, $\beta$, and~$\gamma$ denote the average input data size, the average computing workload, and the average result data size, respectively. Each server can reserve a certain amount of computing resource (in CPU cycles per second) for VMs to execute computing tasks. Let~$c_{e,k}^{n}$ denote the amount of computing resource of server $e$ reserved for group $n$ during time interval~$k$, and define $\bm{c}_k = [c_{e,k}^{n}]_{\forall e \in \mathcal E, n \in \mathcal N} $ as the computing resource reservation decision in time interval~$k$. For time interval~$k$, the computing resource reservation should satisfy the following constraint:
	\begin{equation}\label{eq3}
		\sum_{n \in \mathcal N}{ c_{e,k}^{n} } \leq C_{e}, \,\, \forall e \in \mathcal{E}, k \in \mathcal{K},
	\end{equation}
where $C_{e}$ denotes the maximum computing resource at server~$e$ that can be utilized for the stateful application. We assume that the computing resources on the S-CN, i.e., $e = e^\text{cn}$, is sufficient for executing all computing tasks. The time that server~$e$ takes for executing the computing tasks assigned to it from group~$n$ during time interval~$k$ should satisfy the following requirement:
	\begin{equation}\label{eq4}
		\frac{m_{e,k}^{n} \alpha \beta}{c_{e,k}^{n}} \leq \tau^\text{p}, \,\, \forall e \in \mathcal{E}, n \in \mathcal{N}, k \in \mathcal{K},
	\end{equation}
where $\tau^\text{p}$ denotes the maximum tolerable computation time.\footnote{We do not consider queuing delay in the planning stage because we do not assume a particular computing task arrival pattern. For modeling the queueing delay, the task arrival pattern must be known~\emph{a priori}. However, such a pattern is usually unavailable in practice, while assuming a particular arrival pattern can oversimplify the scenario.} Due to~\eqref{eq4}, computing resource reservation and computing task assignment for each server are mutually dependent.

Let $\epsilon_{e,k}^\text{c}$ represent the overall computing resource usage for executing all computing tasks assigned to server~$e$ in time interval~$k$, which is computing load-dependent. Based on~\cite{dayarathna2015data,chen2018computation}, we adopt a linear model of computing resource usage as follows:
	\begin{equation}\label{eq10}
		 \epsilon_{e,k}^\text{c} = \sum_{n \in \mathcal N}{\varepsilon_{e}^\text{c} m_{e,k}^{n}  }, \,\, \forall e \in \mathcal E, k \in \mathcal{K},
	\end{equation}
where $\varepsilon_{e}^\text{c}$ is the computing resource usage for executing each computing task at server~$e$.

\subsection{Storage Model \& Remote Access}

Different from stateless applications, the execution of a computing task for stateful applications requires the corresponding context data. If the context data is not in the storage of the server, the server should download the context data from a remote server, thereby yielding additional communication resource usage. We model the storage and the additional communication resource usage for the stateful application in this subsection.

Denote the amount of the reserved storage resource (in bits) of server~$e$ during time interval~$k$ and the storage capacity (in bits) of server $e$ by $g_{e, k}$ and $G_{e}$, respectively. The value of $g_{e, k}$ should satisfy the following constraint:
	\begin{equation}\label{eq5}
		g_{e,k} \leq G_{e}, \,\, \forall e \in \mathcal{E} \backslash \{e^\text{cn} \}, n \in \mathcal{N}, k \in \mathcal{K}.
	\end{equation}
We assume that the storage resources reserved on the S-CN, i.e., $e = e^\text{cn}$, is sufficient for storing all context data. Define $\bm{g}_k = [g_{e,k}]_{\forall e \in \mathcal E \backslash \{e^\text{cn} \}} $ as the storage resource reservation decision for all S-NAPs and S-BSs in time interval~$k$. Based on the model of storage resource usage in~\cite{ruiz2012model}, the resource usage for reserving storage resource at server~$e$ in time interval~$k$, denoted by $\epsilon_{e,k}^\text{s}$, is given by 
 	\begin{equation}\label{eq11}
		\epsilon_{e,k}^\text{s} =\varepsilon_{e}^\text{s} g_{e,k}, \,\, \forall e \in \mathcal E, k \in \mathcal{K},
	\end{equation}
where $\varepsilon_{e}^\text{s}$ represents the per bit resource usage for reserving storage resource at server~$e$.

Let $\mathcal{I} = \{1, 2, \cdots, I \}$ represent the set of all chunks of context data in the library for the stateful application, where~$I$ denotes the number of chunks in the set~$\mathcal{I}$. All chunks of context data for the stateful application have the same data structure and thus identical data size (in bits), denoted by~$L$. Executing each computing task requires one chunk of context data in the set~$\mathcal{I}$~\cite{park2019rate}.
Denote the request ratio of chunk~$i$ in the coverage of BS~$b$ in time interval~$k$ by~$p_{b, k}^{i}$, i.e., the load of computing tasks requesting chunk~$i \in \mathcal{I}$ over the load of all computing tasks generated in the coverage of BS~$b$ in time interval~$k$. The value of $p_{b, k}^{i}$ may be different in the coverage of different BSs and may vary across different time intervals. Let $\bm{p}_{k} = [ p_{b, k}^{i} ]_{\forall b \in \mathcal{B}, i \in \mathcal{I}}$ denote the chunk request ratio profile in time interval~$k$, which can be obtained via prediction based on historical data contained in UDTs~\cite{paschos2018role}.

The overall data volume of all chunks may be much larger than the storage capacities of S-BSs and S-NAPs. Each S-BS and S-NAP can only store some chunks of context data for executing computing tasks prior to the beginning of each time interval. The S-CN stores all chunks of context data. Let $\mathcal{I}_{e, k} \subseteq \mathcal{I}$ denote the set of chunks stored at server~$e$ in time interval~$k$. Given the amount of reserved storage resource on server~$e$, i.e.,~$\bm{g}_k$, the number of chunks in set~$\mathcal{I}_{e, k}$ is $|\mathcal{I}_{e, k}| = \lfloor  g_{e}/ L \rfloor$ where $\lfloor \cdot \rfloor$ represents the floor function. 

Since our focus is storage resource reservation, we follow the hierarchical storage policy in~\cite{dai2012collaborative,poularakis2016complexity} to determine the set of stored chunks on each S-BS and S-NAP, i.e.,~$\mathcal{I}_{e, k}$, based on the \emph{effective request ratio} of chunks. Define the effective request ratio of chunk~$i$ for executing the computing tasks assigned to server~$e$ in time interval~$k$ as $q_{e,k}^{i}$, i.e., the load of computing tasks requesting chunk~$i \in \mathcal{I}$ over the load of all computing tasks generated in the service coverage of server~$e$ in time interval~$k$. In multi-tier computing, the stored chunks on S-NAPs and the S-CN may not be requested for computing task executing when the chunks are stored on S-BSs for the sake of reducing communication resource usage. As a result, the effective request ratio of a chunk for an S-NAP depends on not only the chunk request ratio profile, i.e.,~$\bm{p}_{k}$, but also the set of chunks stored on S-BSs and computing task assignment, which is difficult to determine.  

For simplicity, we make two following assumptions to estimate the effective request ratio for each S-NAP. First, S-BS~$e$ located at BS~$b$ stores~$|\mathcal{I}_{e, k}|$ chunks with largest $p_{b, k}^{i}$. Second, given set~$\mathcal{I}_{e, k}$ for any S-BS, the computing tasks requesting a chunk stored on the S-BS are assigned to the S-BS as much as possible when not violating the computing resource capacity of the S-BS~\cite{poularakis2016complexity}. Based on the two assumptions, the value of~$q_{e,k}^{i}$ for an S-BS equals to the request ratio of chunk~$i$ in the coverage of the corresponding BS in time interval~$k$, i.e., $p_{b,k}^{i}$; The value of $q_{e,k}^{i}(\bm{g}_k, \bm{p}_k)$ for an S-NAP can be estimated given~$\bm{g}_k$ and $\bm{p}_k$, which is detailed in Appendix~\ref{app1}. Each S-BS and S-NAP sorts the chunks in a non-increasing order based on the effective request ratio, i.e., $q_{e,k}^{i}(\bm{g}_k, \bm{p}_k)$, and stores the most requested~$|\mathcal{I}_{e, k}|$ chunks. 

When the chunk required for executing a computing task is not found in the storage of an S-BS, the S-BS will first attempt to download the chunk of context data from the S-NAP covering it and resort to the S-CN in the case that the S-NAP does not have the chunk of context data either. When the chunk required for executing a computing task is not found in the storage of an S-NAP, the server will download the context data from the S-CN. For any server, accessing another server remotely and downloading context data from the remote server yields additional communication resource usage, referred to as remote access of context data~\cite{zhang2018survey}.

Denote the data volume (in bits) of remotely accessing context data for each computing task by~$L^\text{re}$, the value of~which may be different from the value of $L$ due to headers used by transmission protocols. Let $f^{n,i}_{b,e,k}$ denote the number of computing tasks that require chunk~$i$ and are generated by group $n$ in the coverage of BS~$b$ and assigned to server~$e$ during time interval~$k$. The relation between $f_{b,e,k}^{n}$ defined in Eq.~\eqref{eq1} and $f_{b,e,k}^{n,i}$ is $f_{b,e,k}^{n} = \sum_{\mathcal{I}}{f_{b,e,k}^{n,i}}$. Define the computing task assignment in time interval~$k$ as $\bm{f}_{k} = [f_{b,e,k}^{n,i} ]_{\forall b \in \mathcal B, \forall e \in \mathcal{E}_{b}, n \in \mathcal N, i \in \mathcal{I}_{e,k}}$. For S-BS~$e$, the communication resource usage for downloading context data from S-NAP~$e'$ covering S-BS~$e$ for executing computing tasks from group~$n$ during time interval~$k$ is given by:

		\begin{equation}\label{eqp6}
			\begin{split}
			v_{n,e,k}^\text{re} = \,\,& L^\text{re} \xi_{e}^{e'} \sum_{i \in \overline{\mathcal{I}_{e,k}} \cap \mathcal{I}_{e',k} }{ \sum_{b \in \mathcal{B}_{e}}{f^{n,i}_{b,e,k}}   } + \\
			\,\,& L^\text{re} \xi_{e}^{e^\text{cn}} \sum_{i \in \overline{\mathcal{I}_{e,k}} \cup \overline{\mathcal{I}_{e',k}}  }{ \sum_{b \in \mathcal{B}_{e}}{f^{n,i}_{b,e,k}}  }, \,\, \forall e \in \mathcal{E}^\text{bs},
			\end{split}
	\end{equation}
where coefficients~$\xi_{e}^{e'}$ and~$\xi_{e}^{e^\text{cn}}$ represent the communication resource usage for downloading per bit context data to S-BS~$e$ from S-NAP~$e'$ and the S-CN, i.e.,~$e^\text{cn}$, respectively. In Eq.~\eqref{eqp6}, the first term represents the communication resource usage for server~$e$ to access the context data remotely from server~$e'$ at the NAP, and the second term represents the communication resource usage for server~$e$ to access the context data remotely from S-CN~$e^\text{cn}$. For S-NAP~$e$, the communication resource usage for downloading context data from the S-CN during time interval~$k$ is as follows:
	\begin{equation}\label{eq7}
		v_{n,e,k}^\text{re} = L^\text{re} \xi_{e}^{e^\text{cn}} \sum_{i \in \overline{\mathcal{I}_{e,k}} }{ \sum_{b \in \mathcal{B}_{e}}{f^{n,i}_{b,e,k}}  }, \,\, \forall e \in \mathcal{E}^\text{nap},
	\end{equation}
where $\xi_{e}^{e^\text{cn}}$ denotes the communication resource usage for downloading per bit context data to S-NAP~$e$ from the S-CN, i.e.,~$e^\text{cn}$. 

\subsection{Communication Model}

Generally, communication resource usage for uploading the input data and downloading the result of a computing task involves two parts: (i) the resource usage for the wireless communication between the UT and the connected BS; and (ii) the resource usage for the wired communication between the BS and a server. In the considered scenario, each UT is associated with only one BS. Regardless of which computer server deployed in the multi-tier network is selected for executing the computing task, the resource consumption for the wireless communication between the associated BS and the UT for uploading input data and downloading computing results is a constant. As a result, the wireless communication does not affect the solution of the resource reservation problem. Since S-BSs and BSs are co-located, there is no additional communication resource usage if any computing task is processed at an S-BS. 

Denote the maximum communication resource usage of server~$e \in \mathcal{E}^\text{nap} \cup \{ e^\text{cn}\}$ for uploading input data and downloading result data by $V_{e}^\text{up}$ and $V_{e}^\text{down}$, respectively. Let~$\eta_{b,e}$ denote the coefficient representing the communication resource usage for uploading and downloading per bit data between BS~$b$ and server~$e$. The communication resource usage for uploading input data and downloading computing results between server~$e$ and BS~$b$ during time interval $k$ are given by:
	\begin{equation}\label{eq8}
		v_{b, e, k}^\text{up} = \alpha \eta_{b,e} \sum_{n \in \mathcal{N}}{\sum_{i \in \mathcal{I}}{f_{b,e,k}^{n,i}}}, \,\, \forall e \in \mathcal{E} \backslash \{e^\text{bs} \}, b \in \mathcal{B}_{e}, k \in \mathcal{K},
	\end{equation}
and
	\begin{equation}\label{eq9}
		v_{b, e, k}^\text{down} = \gamma \eta_{b,e} \sum_{n \in \mathcal{N}}{\sum_{i \in \mathcal{I}}{f_{b,e,k}^{n,i}}}, \,\, \forall e \in \mathcal{E} \backslash \{e^\text{bs} \}, b \in \mathcal{B}_{e}, k \in \mathcal{K},
	\end{equation}
respectively. 

\section{Problem Formulation}

We formulate the problem of planning-stage resource reservation for multi-tier computing to support stateful applications in this section. We aim to find out the minimum amount of network resources needed for supporting the application while balancing the resource usage and the cost from reconfiguring resource reservation in the presence of network dynamics.

Let $\bm{r}_k = [\bm{c}_k, \bm{g}_k, \bm{f}_k]$ denote the resource reservation decision, including computing resource reservation, storage resource reservation, and computing task assignment, in time interval~$k$. Variable~$\bm{r}_k$ is determined at the beginning of time interval~$k$. Given $\bm{r}_k$, the overall network resource usage in time interval~$k$, denoted by $\Delta_{k}(\bm{r}_k)$, can be obtained based on Eqs.~\eqref{eq10},~\eqref{eq11},~\eqref{eq8}, and~\eqref{eq9}, as follows:
 	\begin{equation}\label{eq12}
		\Delta_{k} (\bm{r}_k) = \sum_{e \in \mathcal{E}}{w^\text{c} \epsilon_{e,k}^\text{c} +  w^\text{s}\epsilon_{e,k}^\text{s} + w^\text{o}(v_{e, k}^\text{re} + v^\text{co}_{e,k})},
	\end{equation}
where $v^\text{co}_{e,k} = \sum_{b \in \mathcal B}{(v_{b, e, k}^\text{up} + v_{b, e, k}^\text{down})}$, and $w^\text{c}$, $w^\text{s}$, and $w^\text{o}$ are the weights of the computing, storage, and communication resource usage, respectively. Since the spatial task distribution may vary across time intervals, even if a resource reservation decision can minimize instantaneous resource usage in Eq.~\eqref{eq12} in time interval~$k$, the same decision may not minimize the overall network resource usage in the subsequent time intervals. Reconfiguration is required for the resource reservation to adapt to the dynamic spatial task distribution, while reconfiguring resource reservation yields additional cost, e.g., the cost from vertical scaling of VM~\cite{hwang2014scale}. Denote by $O^\text{v}$ the cost from reconfiguring resource reservation. 

Let $a_{k} \in \{0, 1\}$ indicate whether the resource reservation in time interval~$k$ should be reconfigured or not, which is determined at the beginning of time interval~$k$. If $a_{k} = 0$, the controller makes new resource reservation decision for time interval~$k$; Otherwise, the controller keeps using the resource reservation from time interval~$k-1$. Define $\tilde{\bm{x}}_{k} = [\tilde{x}_{b, k}^{n} ]_{\forall b \in \mathcal{B}, n \in \mathcal{N}}$ as spatial task distributions of groups, referred to as group-based spatial task distribution. Let function $\bm{r}_k = \psi(\tilde{\bm{x}}_{k}, \bm{p}_{k})$, representing that the resource reservation decision is made according to the group-based spatial task distribution and the chunk request profile in time interval~$k$. The value of $r_{k}$, resource reservation in time interval~$k$, evolves as follows:
	\begin{equation}\label{eq13}
		\bm{r}_{k} = (1-a_{k}) \psi(\tilde{\bm{x}}_k, \bm{p}_{k}) + a_{k}\bm{r}_{k-1}, \,\, \forall k \in \mathcal K.
	\end{equation}
Given the value of $a_{k}$, the cost from reconfiguring resource reservation in time interval~$k$, denoted by $o^\text{v}_{k}$, is as follows:
	\begin{equation}\label{eq14}
		o^\text{v}_{k} = (1-a_{k}) O^\text{v}, \,\, \forall k \in \mathcal K. 
	\end{equation}

The problem of minimizing the long-term network resource usage and the cost from reconfiguring resource reservation over $K$ time intervals, is formulated as follows: 
	\begin{subequations}\label{p0}
		\begin{align}
			\textrm{P0:} \min_{ \bm{a}, \bm{R} } \,\,&  \sum_{k \in \mathcal K}{ \frac{ \Delta_{k} (\bm{r}_k) + \lambda o^\text{v}_{k} }{  \sum_{b \in \mathcal B}{\sum_{n \in \mathcal N}{\tilde{x}_{b, k}^{n} }} }  } \\ 
			\textrm{s.t.} \,\,& \eqref{eq2},  \eqref{eq3}, \eqref{eq4}, \eqref{eq5}\\ 
			\,\, & \sum_{b \in \mathcal{B}_{e}}{v_{b, e, k}^\text{up}} \leq V_{e}^\text{up}, \,\, \forall e \in \mathcal{E} \setminus \mathcal{E}^\text{bs}, k \in \mathcal{K},\\
			\,\, &  \sum_{b \in \mathcal{B}_{e}}{v_{b, e, k}^\text{down}} \leq V_{e}^\text{down}, \,\, \forall e \in \mathcal{E} \setminus \mathcal{E}^\text{bs}, k \in \mathcal{K},\\
			\,\, & \sum_{e \in \mathcal{E}}{  v_{n, e, k}^\text{re} } \leq V^\text{re}_{n}, n \in \mathcal N, k \in \mathcal K,\\
			\,\, & c_{e,k}^{n} \in \mathbb{R}^{+},\,\,  \forall e \in \mathcal{E}, n \in \mathcal N, k \in \mathcal K,\\
			\,\, & g_{e,k} \in \mathbb{R}^{+},\,\,  \forall e \in \mathcal{E}  \backslash \{e^\text{cn} \}, n \in \mathcal N, k \in \mathcal K,\\
			\,\, & f_{b,e,k}^{n, i} \in \mathbb{N},\,\,  \forall b \in \mathcal{B}, e \in \mathcal{E}, n \in \mathcal N, k \in \mathcal K,\\
			\,\, & a_{k} \in \{0, 1\},\,\, \forall k \in \mathcal{K},
		\end{align}	
	\end{subequations}
where $V^\text{re}$ denotes the maximum communication resource usage of remote access for one computing task, $\bm{a} = [a_{k}]_{\forall k \in \mathcal{K}}$, $\bm{R} = [\bm{r}_{k}]_{\forall k \in \mathcal{K}}$, and $\lambda$ is the weight balancing the network resource usage and the cost from reconfiguring resource reservation. $\mathbb{R}^{+}$ represents the set of positive real numbers, and $\mathbb{N}$ represents the set of natural numbers. Constraint~(\ref{p0}e) limits the communication resource usage of remote access averaged over all computing tasks from each group to be less than $V^\text{re}_{n}$. The solution of \eqref{p0} provide a lower bound on the resources needed to support the stateful application, taking into account the resource reservation reconfiguration cost in network planning. If the network resources are sufficient, more resources can be reserved for the application for better service quality. In addition, in a practical network supporting multiple applications, statistical multiplexing among resources reserved for different applications can be implemented for high resource utilization.

Solving Problem~$\textrm{P0}$ is challenging due to the following two reasons: (i) For each time interval, determining the value of $\bm{r}_{k}$ is a mix-integer optimization problem, and the variables of $\bm{f}_{k}$ and $\bm{g}_{k}$ are mutually dependent; (ii) determining the value of $a_{k}$ is a sequential decision making problem, and the decision at any time interval affects the subsequent decisions. To solve Problem~P0, we decouple it into two problems. We propose algorithms for resource reservation in each time interval, which are presented in Section~V, and a learning-based approach for reconfiguring resource reservation over multiple time intervals, which is presented in Section~VI.

\section{Group-based Resource Reservation}

In this section, we design an algorithm to enable group-based resource reservation in each time interval. 

When $a_{k} = 1$, the controller keeps using the resource reservation decision from time interval~$k-1$. For when $a_{k} = 0$, we formulate the group-based multi-resource reservation problem in time interval $k$, given the predicted spatial task distributions~$\tilde{\bm{x}}_{k}$, as follows:
	\begin{equation}\label{p1}
		\begin{split}
			\textrm{P1:} \min_{ \bm{r}_{k} } \,\,& \frac{ \Delta_{k} (\bm{r}_k) }{ \sum_{b \in \mathcal B}{\sum_{n \in \mathcal N}{\tilde{x}_{b, k}^{n}}}  }  \\ 
			\textrm{s.t.} \,\,& \eqref{eq2}, \eqref{eq3}, \eqref{eq4}, \eqref{eq5}, \text{(\ref{p0}c-h)}. 
		\end{split}	
	\end{equation}
Problem~P1 is a combinatorial optimization problem, and variables $\bm{f}_{k}$ and $\bm{g}_{k}$ are still coupled. We first address computing task assignment, i.e.,~$\bm{f}_{k}$, and computing resource reservation, i.e.,~$\bm{c}_{k}$, given a storage resource reservation decision, i.e., $\bm{g}_{k}$. Then, we leverage particle swarm optimization to find the solution of storage resource reservation. The solution of Problem~P1 corresponds to group-specific resource reservation, and the total amount resources to be reserved for all groups can be calculated accordingly. The reserved resources can be multiplexed among different groups.

\subsection{Computing Task Assignment and Computing Resource Reservation}

Since multiple computing tasks can be assigned to the same server, assigning computing tasks to servers is many-to-one matching. We first transform the many-to-one matching into a one-to-one matching. Specifically, several \emph{virtual servers} are created to represent each physical server. Denote the number of virtual servers for S-NAP or S-BS~$e$ by $N_{e}$. The value of $N_{e}$ is the maximum number of computing tasks that can be assigned to server~$e$ while not violating the constraints of resource usage, i.e., constraints~\eqref{eq3},~(\ref{p0}c), and~(\ref{p0}d). The number of virtual servers for physical server~$e$ is given by:
	\begin{equation}\label{eq17r}
		\begin{aligned}
			& N_{e} = \\
			& \left\{ \begin{array}{l}
			 	\min \{ \lfloor \frac{\tau^\text{p} C_{e}}{\alpha \beta}  \rfloor, \lfloor \frac{V^\text{up}_{e}}{ \sum_{\mathcal{B}_{e}}{v_{b,e,k}^\text{up} }} \rfloor,  \lfloor \frac{V^\text{down}_{e}}{ \sum_{\mathcal{B}_{e}}{v_{b,e,k}^\text{down} } }  \rfloor \},\,\, \text{if}\,\, e \in \mathcal{E}^\text{nap};\\ 
				 \lfloor \frac{\tau^\text{p} C_{e}}{\alpha \beta}  \rfloor,\,\,  \text{if}\,\, e \in \mathcal{E}^\text{bs},
			\end{array} \right.
		\end{aligned}
	\end{equation}
where $ \lfloor \cdot \rfloor$ represents the floor function. For an S-NAP, i.e., $e \in \mathcal{E}^\text{nap}$, the value of $N_{e}$ in Eq.~\eqref{eq17r} is the minimum value among the maximum number of computing tasks that can be executed, i.e., $\lfloor \frac{\tau^\text{p} C_{e}}{\alpha \beta}  \rfloor$, the maximum number of computing tasks that can be uploaded, i.e., $\lfloor \frac{V^\text{up}_{e}}{ \sum_{\mathcal{B}_{e}}{v_{b,e,k}^\text{up} }} \rfloor$, and the maximum number of computing tasks that can be downloaded, i.e., $\lfloor \frac{V^\text{down}_{e}}{ \sum_{\mathcal{B}_{e}}{v_{b,e,k}^\text{down} } }  \rfloor$; and for an S-BS, i.e., $e \in \mathcal{E}^\text{bs}$, the value of $N_{e}$ in Eq.~\eqref{eq17r} is the maximum number of computing tasks that can be executed, i.e.,~$\lfloor \frac{\tau^\text{p} C_{e}}{\alpha \beta}  \rfloor$. We let the number of the corresponding virtual servers for the S-CN in time interval~$k$ be $\sum_{b \in \mathcal{B}}{\sum_{n \in \mathcal{N}}{\tilde{x}_{b, k}^{n}}}$, i.e., the load of all computing tasks in time interval~$k$, to guarantee that all computing tasks can be processed. Each virtual server is assigned at most one computing task, and assigning computing tasks to virtual servers becomes one-to-one matching. 

Given the amount of storage resource reserved on S-BSs and S-NAPs, the sets of stored context data chunks, i.e.,~$\mathcal{I}_{e,k}$, on all S-BSs and S-NAPs are determined based on the hierarchal storage policy described in Section~III.D. Denote by $D^{n, i}_{b, e, k}$ the network resource usage, including resource usage from computing, uplink communication, downlink communication, and remote access, for executing a computing task that requests chunk~$i$ and is generated by a UT from group~$n$ in the coverage of BS~$b$ during time interval~$k$ at server~$e$. The calculation of $D^{n, i}_{b, e, k}$ consists of two parts. The first part is the resource usage from computing, uplink communication, and downlink communication, which is not related to the sets of stored chunks, while the second part is the communication resource usage from remote access, which depends on the sets of stored chunks. Denote by $W_{b,e} = w^\text{c} \varepsilon^\text{c}_{e} + w^\text{o} (\alpha + \gamma) \eta_{b,e}$ the sum of resource usage from computing, uplink communication, and downlink communication used to execute a computing task generated in the coverage of BS~$b$ at server~$e$. According to the communication resource usage for remote access, the calculation of $D^{n, i}_{b, e, k}$ is categorized into the following four cases:
	\begin{equation}\label{eq19p}
		\begin{aligned}
			& D^{n, i}_{b, e, k} = \\
			& \left\{ \begin{array}{l}
			 	  
				 W_{b, e} + w^\text{o} L^\text{re} \xi_{e}^{e'}, \,\, \text{if}\,\, e \in \mathcal{E}^\text{bs}, e' \in \mathcal{E}^\text{nap}, i \in \overline{\mathcal{I}_{e,k}} \cap \mathcal{I}_{e',k};\\
				 W_{b, e} + w^\text{o} L^\text{re} \xi_{e}^{e^\text{cn}}, \,\, \text{if}\,\, e \in \mathcal{E}^\text{bs}, e' \in \mathcal{E}^\text{nap}, i \in \overline{\mathcal{I}_{e,k}} \cup \overline{\mathcal{I}_{e',k}};\\
				 W_{b, e} + w^\text{o} L^\text{re} \xi_{e}^{e^\text{cn}}, \,\, \text{if}\,\, e \in \mathcal{E}^\text{nap}, i \not \in \mathcal{I}_{e, k};\\
				 W_{b, e}, \,\, \text{otherwise}.
			\end{array} \right.
		\end{aligned}
	\end{equation}
In Eq.~\eqref{eq19p}, if chunk~$i$ is not stored on S-BS~$e$ but stored on S-NAP~$e'$, i.e., $i \in \overline{\mathcal{I}_{e,k}} \cap \mathcal{I}_{e',k}$, the communication resource usage for S-BS $e$ to remotely access S-NAP~$e'$ for one computing task is $w^\text{o} L^\text{re} \xi_{e}^{e'}$; If chunk~$i$ is not stored on S-BS~$e$ or S-NAP~$e'$, i.e., $i \in \overline{\mathcal{I}_{e,k}} \cup \overline{\mathcal{I}_{e',k}}$, the communication resource usage for S-BS $e$ to remotely access the S-CN for one computing task is $w^\text{o} L^\text{re} \xi_{e}^{e^\text{cn}}$; If chunk~$i$ is not stored on S-NAP~$e$, i.e., $i \not \in \mathcal{I}_{e, k}$, the communication resource usage for S-NAP $e$ to remotely access the S-CN for one computing task is $w^\text{o} L^\text{re} \xi_{e}^{e^\text{cn}}$; Otherwise, chunk~$i$ is stored on server~$e$, and no communication resource is used for remote access.

	\begin{algorithm}[t] 
		\caption{MCLA Algorithm}\label{alg1}
		\LinesNumbered
		\textbf{Input:} $\mathcal{L}^\text{server}_{u, k}, \forall u \in \mathcal{U}$\\
		\textbf{Initialization:} $\mathcal{U}$, $\mathcal{U}^\text{not} = \mathcal{U}$, $\mathcal{T}$, $j=0$;\\
		\While{$|\mathcal{T}| = 0$}
		{	
			$j = j + 1$;\\
		    \For{$u \in \mathcal{U}^\text{not}$}
			    {	
			     	Select the first computing task in preference list~$\mathcal{L}^\text{server}_{u, k}$ as the proposal from virtual server~$u$;\\
			     	Remove the selected computing task from preference list~$\mathcal{L}^\text{server}_{u, k}$;\\  
			    }

		     	Adopt the dynamic programming in~\cite{kellerer2004multidimensional} to select proposals from the new proposals in iteration~$j$ and adjust the matched proposals in iteration~$j-1$ for minimizing the objective function in \eqref{p1p} while satisfying constraint~(\ref{p0}e). 

		     	Remove the matched virtual servers from $\mathcal{U}^\text{not}$;\\
		 		Add the unmatched virtual servers to $\mathcal{U}^\text{not}$;\\
		 		Remove the matched computing tasks from $\mathcal{T}$ based on the matching result in iteration~$j$;\\ 

		}
		$\bm{f}_{k} \leftarrow \bm{z}_{k}$;\\ 
	 \textbf{Output:} $\bm{f}_{k}$
	\end{algorithm} 

Denote the set of virtual servers and the set of computing tasks by $\mathcal{U}$ and $\mathcal{T}$, respectively. Given computing task~$t \in \mathcal{T}$ and virtual server~$u \in \mathcal{U}$, we can determine the corresponding values of $(n,i,b)$ of the computing task and physical server~$e$. Let $D^{(u,t)}_{k}$ represent the sum of computing and communication resource usage for executing computing task~$t$ at virtual server~$u$ in time interval~$k$, which can be calculated via Eq.~\eqref{eq19p} based on the corresponding values of $(n,i,b)$ and physical server~$e$. We introduce variable $z^{(u,t)}_{k} \in \left\{0, 1\right\}$ to indicate whether to assign computing task~$t$ to virtual server~$u$ in time interval~$k$ or not and define $\bm{z}_{k} = [ z^{(u,t)}_{k} ]_{\forall u \in \mathcal{U},t \in \mathcal{T}}$. If computing task~$t$ is assigned to virtual server~$u$, $z^{(u,t)}_{k} = 1$; Otherwise, $z^{(u,t)}_{k} = 0$. Finding the solution of $\bm{f}_{k}$ in Problem~P1 can be transformed into finding the solution of $\bm{z}_{k}$, as follows:
	\begin{equation}\label{p1p}
		\begin{split}
			\textrm{P2:} \min_{ \bm{z}_{k} } \,\,& \sum_{u \in \mathcal{U}}{\sum_{t \in \mathcal{T}}{D^{(u,t)}_{k}  z^{(u,t)}_{k}}}  \\ 
			\textrm{s.t.} \,\,&  \text{(\ref{p0}e)},\\
			\,\, & z^{(u,t)}_{k} \in \{0, 1\},\,\, \forall k \in \mathcal{K}.
		\end{split}	
	\end{equation}
We propose a matching-based computing task assignment (MCLA) algorithm to select a virtual server for each computing task, as shown in Algorithm~\ref{alg1}.

We construct the preference list~$\mathcal{L}^\text{server}_{u, k}$ for virtual server~$u$ in time interval~$k$, which is a vector containing the indexes of all computing tasks that can be assigned to virtual server~$u$ in time interval~$k$, sorted by the value of $D^{(u,t)}_{k}$ in a non-decreasing order. In each iteration, the controller checks the current preference list for each virtual server and selects the first computing task in the preference list as the proposal from the virtual server. After that, the selected computing task is removed from the virtual server's preference list. The controller selects a proposal for each computing task to minimize the objective function in Problem~P2 while satisfying constraint~(\ref{p0}e), which is a 0-1 Knapsack problem. Let~$\mathcal{U}^\text{not} \subset \mathcal{U}$ denote the set of virtual servers that are not yet matched. A dynamic programming approach in~\cite{kellerer2004multidimensional} is adopted to adjust the matching result, i.e., selecting proposals for virtual servers in the set~$\mathcal{U}^\text{not}$ and adjusting the matched proposals for virtual servers in the set~$\mathcal{U} \setminus \mathcal{U}^\text{not}$ for minimizing the objective function in Problem~P2 while satisfying constraint~(\ref{p0}e). Then, set~$\mathcal{U}^\text{not}$ is updated accordingly after each iteration. A computing task should be re-proposed for each virtual server in the set~$\mathcal{U}^\text{not}$ in the next iteration. The matching process terminates when all computing tasks in the set~$\mathcal{T}$ are successfully matched. Based on the matching result, i.e., $\bm{z}_{k}$, we can determine the computing task assignment decision, i.e., $\bm{f}_{k}$.

Given $\bm{f}_k$, the numbers of computing tasks assigned to the servers, i.e., $\bm{m}_k$, are determined based on Eq.~\eqref{eq2}. Then, computing resource reservation in time interval $k$ is determined according to Eq.~\eqref{eq3}, which is:

	\begin{equation}\label{eq20p}
		\bm{c}_k = \frac{\alpha \beta}{\tau^\text{p}} \bm{m}_k, \,\, \forall k \in \mathcal{K}.
	\end{equation}

The time complexity of Algorithm~\ref{alg1} depends on the number of iterations of the outer loop for matching (Lines 5 - 13) and the time complexity of solving the knapsack problems in each iteration (Line 9). For the outer loop, the time complexity of solving the matching problem is $O(Z X_{k})$, where $Z = \sum_{e \in \mathcal{E}}{N_{e}}$ and $X_{k} = \sum_{n \in \mathcal{N}}{\sum_{b \in \mathcal{B}}{x_{b,k}^{n}}}$ denote the number of all virtual servers and the number of all computing tasks in time interval~$k$, respectively~\cite{lovasz2009matching}. In each iteration, we should solve a knapsack problem to satisfy constraint~(\ref{p0}e) for every group. The time complexity of the adopted dynamic programming approach for group~$n$ is $O(V^\text{re}_{n} X_{k}^{n})$, where $X_{k}^{n} = \sum_{b \in \mathcal{B}}{x_{b,k}^{n}}$ denotes the number of all computing tasks from group $n$ in time interval~$k$~\cite{kellerer2004multidimensional}. Therefore, the time complexity of Algorithm~1 is $O(Z X_{k} \sum_{n \in \mathcal{N}}{V^\text{re}_{n} X^{n}_{k}})$.

\subsection{Storage Resource Reservation}

	\begin{algorithm}[t] 
		\caption{RR Algorithm}\label{alg2}
		\LinesNumbered
		\textbf{Input:} $\varsigma$, $\varsigma_{1}$, $\varsigma_{2}$, $\varphi_{1}$, $\varphi_{2}$, $l^\text{max}$\\
		\textbf{Initialization:} $l = 0$, $\bm{s}_{y}^{(0)} = \bm{0}$, $\bm{g}_{y}^{(0)} \in \mathbb{F}$, $\forall y \in \mathcal{Y}$;\\
		$\bm{f}_{y}^{(0)}, \forall y \in \mathcal{Y} \leftarrow$ Calculate by Algorithm~\ref{alg1} given $\bm{g}_{y}^{(0)}$;\\
		$\hat{\Delta}_{y}, \forall y \in \mathcal{Y} \leftarrow$ Calculate by \eqref{eq12} given $\bm{c}_{y}^{(0)}$, $\bm{g}_{y}^{(0)}$, and $\bm{f}_{y}^{(0)}$;\\
		$\hat{\bm{g}}_{y}, \forall y \in \mathcal{Y} \leftarrow \bm{g}_{y}^{(0)}, \forall y \in \mathcal{Y}$;\\
		$\Delta^{*} \leftarrow \min \left\{ \hat{\Delta}_{y}, \forall y \in \mathcal{Y} \right\}$;\\
		$\bm{g}^{*} \leftarrow \hat{\bm{g}}_{y'}$ where $y' = \arg \min_{y} \left\{ \hat{\Delta}_{y}, \forall y \in \mathcal{Y} \right\}$;\\

		\While{$l \leq l^\text{max}$}
		{	
		    \For{$y \in \mathcal{Y}$}
			    {	
			    	\If{ constraint~\eqref{eq5} is not satisfied}
			    		{
			    		$\bm{g}_{y}^{(l)} \leftarrow$ Select a position in $\mathbb{F}$ randomly;\\
			    		}
					$\bm{f}_{y}^{(l)} \leftarrow$ Calculate by Algorithm~\ref{alg1} given $\bm{g}_{y}^{(l)}$;\\
					$\Delta_{y}^{(l)} \leftarrow $ Calculate by Eq.~\eqref{eq12} given $\bm{c}_{y}^{(l)}$, $\bm{g}_{y}^{(l)}$, and $\bm{f}_{y}^{(l)}$;\\
					\If{$\Delta_{y}^{(l)} < \hat{\Delta}_{y}$}
						{
							$\hat{\bm{g}}, \hat{\Delta}_{y} \leftarrow \bm{g}^{(l)}_{y}, \Delta_{y}^{(l)}$;\\
						}
					\If{$\Delta_{y}^{(l)} < \Delta^{*}$}
						{
							$\bm{g}^{*}, \Delta^{*} \leftarrow \bm{g}^{(l)}_{y}, \Delta_{y}^{(l)}$;\\
						}
					$\bm{s}_{y}^{(l+1)}, \bm{g}_{y}^{(l+1)} \leftarrow$ Update by Eqs.~\eqref{eq20} and~\eqref{eq21}, respectively;\\
			    }
		 }
		 $\bm{f}^{*}, \bm{c}^{*} \leftarrow$ Calculate by Algorithm~\ref{alg1} and Eq.~\eqref{eq20p} given $\bm{g}^{*}$, respectively;\\
	 \textbf{Output:} $\bm{g}^{*}$, $\bm{f}^{*}$, $\bm{c}^{*}$, $\Delta^{*}$
	\end{algorithm}

Determining the amount of storage resources reserved on S-NAPs and S-BSs is a combinatorial optimization problem. Therefore, finding the globally optimal solution is challenging~\cite{cygan2009exponential}. Evolutionary heuristics, specifically particle swarm optimization (PSO), is leveraged to achieve the local optima of the problem. 
Based on PSO, we propose a resource reservation (RR) algorithm to solve Problem~P1. In the proposed RR algorithm, we leverage a number of particles, referred to as the particle swarm, where the position of each particle corresponds to the solution of storage resource reservation, i.e.,~$\bm{g}_k$. In each iteration, each particle moves within the solution space while adjusting its position and speed dynamically based on~\cite{cygan2009exponential}. After repeating such an iteration multiple times, the positions of all particles can converge to the same position, which is the found solution of storage resource reservation~\cite{clerc2002particle}. Given the storage resource reservation, the values of $\bm{f}_k$ and $\bm{c}_k$ can be determined based on Section~IV.A.  

The detailed procedures of the RR algorithm are introduced in Algorithm~\ref{alg2}. Since the RR algorithm applies to any time interval, we omit subscript~``$k$'' in the rest of this subsection. We define the solution space of the storage resource reservation problem as $\mathbb{F}$, where the possible solution of $\bm{g}_{k}$, i.e., particles' positions, should satisfy constraint~\eqref{eq5}. Denote the set of particles and the position of particle~$y$ in the $l$\,th iteration by $\mathcal{Y}$ and $\bm{g}_{y}^{(l)}$, respectively. Let $\hat{\bm{g}}_{y}$ and~$\bm{g}^{*}$ denote the best position of particle~$y$ and the best position among all particles' positions, i.e., the particle swarm's best position, up to the $l$\,th iteration, respectively. Accordingly, let $\Delta_{y}^{(l)}$, $\hat{\Delta}_{y}$, and~$\Delta^{*}$ denote the value of $\Delta$ in Problem~P1 given position $\bm{g}_{y}^{(l)}$, $\hat{\bm{g}}_{y}$, and~$\bm{g}^{*}$, respectively. Based on~\cite{clerc2002particle}, the speed of particle~$y \in \mathcal{Y}$ in the $l$\,th iteration, denoted by $\bm{s}_{y}^{(l)}$, evolves as follows:
	\begin{equation}\label{eq20}
		\begin{split}
			\bm{s}_{y}^{(l)} = \,\,& \varsigma \bm{s}_{y}^{(l-1)} + \varsigma_{1} \varphi_{1} \left(\hat{\bm{g}}_{y} - \bm{g}_{y}^{(l-1)} \right)  \\
			\,\,& +  \varsigma_{2} \varphi_{2} \left(\bm{g}^{*} - \bm{g}_{y}^{(l-1)} \right), 
		\end{split}
	\end{equation}
where parameter $\varsigma$ is the weight for each particle to keep its speed from the previous iteration. Parameters~$\varsigma_{1}$ and~$\varsigma_{2}$ are cognitive and social coefficients for learning from each particle's own best position and the particle swarm's best position up to the current iteration, respectively, and both are positive random variables for exploring the solution space~\cite{clerc2002particle}. The position of particle~$y$, i.e., $\bm{g}_{y}^{(l)}$, in the $l$\,th iteration is given by:
	\begin{equation}\label{eq21}
		\bm{g}_{y}^{(l)} = \bm{g}_{y}^{(l-1)} + \bm{s}_{y}^{(l)}. 
	\end{equation}
If a particle moves out of the solution space, the particle is replaced by a new particle with a random position in the solution space~$\mathbb{F}$. In this way, the positions of all particles are guaranteed to satisfy constraint~\eqref{eq5}.

Algorithm~\ref{alg2} shows the detail of the proposed RR algorithm. Denote the maximum number of iteration by $l^\text{max}$. Line~2 initializes all particles with the positions in the solution space~$\mathbb{F}$. Line~3 to Line~4 obtain the value of $\Delta$ in Problem~P1, i.e., $\hat{\Delta}_{y}$, given the position of particle~$y$, i.e.,~$\bm{g}_{y}^{(l)}$. Line~5 to Line~7 find the best solution among all particles based on the value of $\hat{\Delta}_{y}$. Line~10 to Line~21 update the position of each particle based on Eqs.~\eqref{eq20} and~\eqref{eq21} in each iteration. The outputs of the RR algorithm include the solution of Problem~P1, i.e., $\bm{g}^{*}$, $\bm{f}^{*}$, and $\bm{c}^{*}$, and the value of $\Delta^{*}$.

\section{Meta Learning based Resource Reservation Reconfiguration}

In the preceding section, we solve the network resource reservation problem for one time interval. In this section, we determine the value of $\bm{a} = [a_{k}]_{\forall k \in \mathcal{K}}$ to reconfigure resource reservation decisions among $K$ time intervals.

We formulate the sub-problem of resource reservation reconfiguration based on Problem~P0, as follows:
	\begin{subequations}\label{p3}
		\begin{align}
			\textrm{P3:} \min_{ \bm{a}} \,\,&  \sum_{k \in \mathcal K}{ \frac{ \Delta_{k} (\bm{r}_k) + \lambda o^\text{v}_{k} }{  \sum_{b \in \mathcal B}{\sum_{n \in \mathcal N}{\tilde{x}_{b, k}^{n} }} } } \\ 
			\textrm{s.t.} \,\,& \text{(\ref{p0}i)}.
		\end{align}	
	\end{subequations}
We define $k^{\star}$ as the time interval when the latest resource reservation was reconfigured up to time interval~$k$. In time interval~$k^{\star}$, $a_{k^{\star}} = 0$, $a_{j}= 1$ for $j \in [k^{\star} +1, \cdots, k-1 ]$ and $k^{\star} < k$. The relation between~$k$ and~$k^{\star}$ is given by:
	\begin{equation}\label{eq25}
	k^{\star} = 
	\left\{
             \begin{array}{lr}
            	k, \,\, \text{if} \,\, a_{k} = 0; &  \\
             	k^{\star}, \,\, \text{otherwise}.&  
             \end{array}
	\right.
	\end{equation}
Problem~P3 is a sequential decision making problem, which can be solved by reinforcement learning (RL) based methods~\cite{zhang2021dynamic,zhou2020deep}. However, RL-based methods cannot be applied directly to solve Problem~P3. Resource reservation reconfiguration is owing to the difference of \emph{network status}, i.e., spatial task distributions and chunk request ratio profiles in this work, in different time intervals. Identifying the difference between network status can potentially improve the learning efficiency of RL-based methods. 

We propose a \underline{Meta}-learning-based \underline{r}esource \underline{r}eservation \underline{r}e-configuration (Meta$\text{R}^{3}$) approach to solve Problem~P3, as given in Algorithm~\ref{alg3}. At the end of each time interval, data regarding spatial task distributions, chunk request ratio profiles, indicator $a_{k}$ and the network performance is collected and stored based on UDTs. Meta learning is adopted to capture the \emph{similarity} between network status in two time intervals~$k$ and~$k^{\star}$, and the policy of resource reservation decision reconfiguration is learned based on the captured similarity by using RL. Based on the learned policy of resource reservation decision reconfiguration, the value of $a_{k}$ can be determined. If $a_{k} = 0$, resource reservation decision is reconfigured at the beginning of time interval~$k$. 

Two components are underlying the proposed Meta$\text{R}^{3}$ approach: (i) capturing the similarity between network status during two different time intervals, and (ii) reconfiguring resource reservation in a closed-loop manner, which are presented in the following two subsections, respectively.

	\begin{figure}[t]
		\centering
	  	\includegraphics[width=0.80\textwidth]{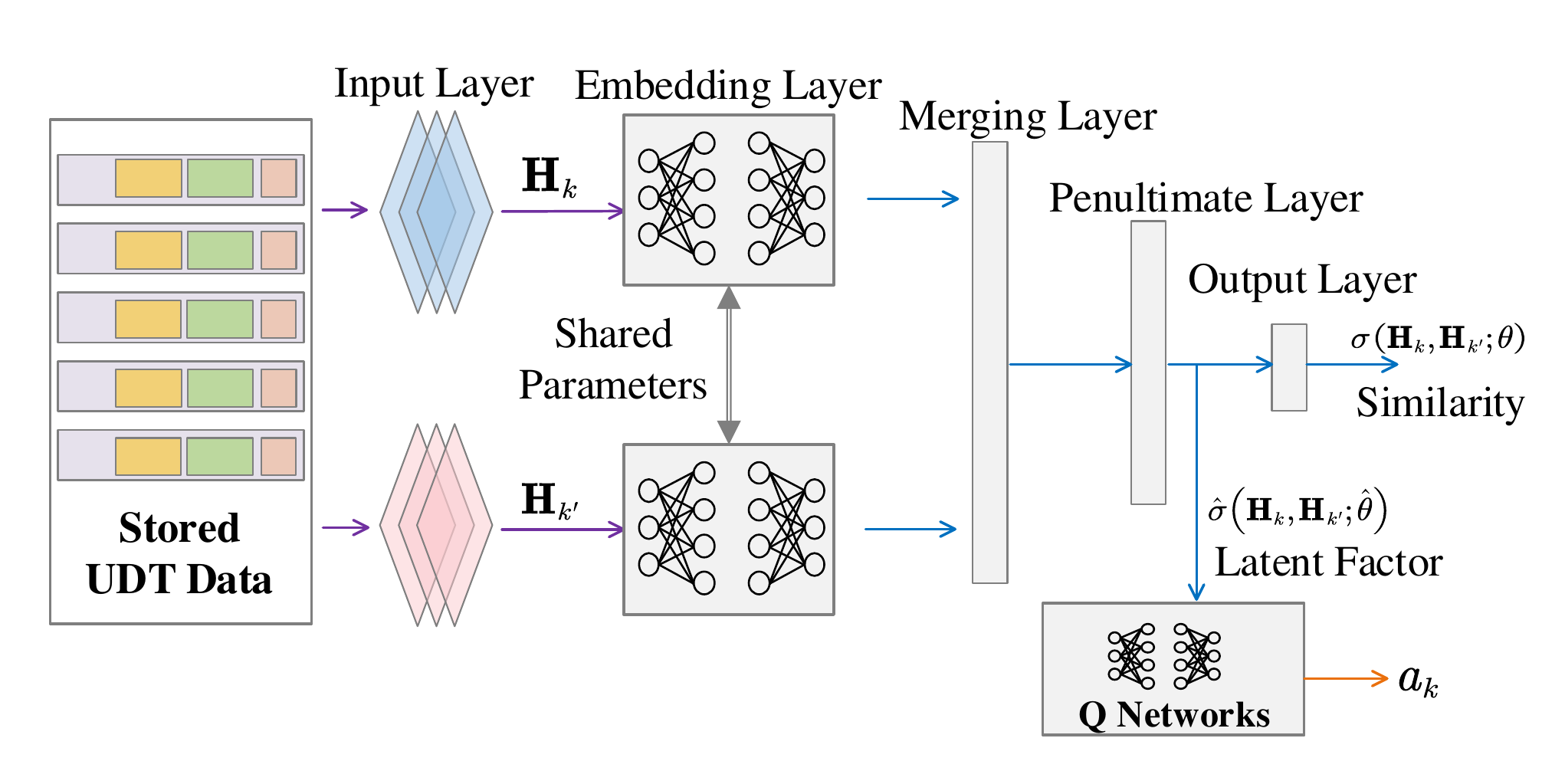}
	  	\caption{The proposed Meta$\text{R}^{3}$ approach.}\label{nn}
	\end{figure}

\subsection{Similarity Capture}

Denote the network status in time interval~$k$ by $\bm{h}_{k} = [\bm{x}_{k}, \bm{p}_{k}]$. The value of $\bm{h}_{k}$ is unavailable at the beginning of time interval~$k$. Therefore, we use the data regarding spatial task distributions and chunk request ratio profiles in past $T'$ time intervals contained in UDTs as the features of the network status in time interval~$k$, denoted by $\bm{H}_{k} = \left[\bm{h}_{k-T'}, \cdots, \bm{h}_{k-1} \right]$. Based on $\bm{H}_{k}$, the value of $a_{k}$ is determined. If $a_{k} = 0$, $\bm{H}_{k}$ can be used to predict the network status, i.e.,~$\bm{h}_{k}$, for making resource reservation decision. Define the similarity between the network status in different time intervals~$k$ and~$k'$ as $\sigma(\bm{H}_{k}, \bm{H}_{k'})$.

We leverage siamese neural networks to approximate the value of $\sigma(\bm{H}_{k}, \bm{H}_{k'})$ \cite{koch2015siamese}, as illustrated in Fig.~\ref{nn}. Let $\boldsymbol{\theta}$ and $\hat{\boldsymbol{\theta}}$ denote the parameters of the whole siamese neural networks and the parameters of the siamese neural networks without the output layer, respectively. The inputs of the siamese neural networks are the features of network status, i.e., $\bm{H}_{k}$ and $\bm{H}_{k'}$. The siamese neural networks have two outputs, i.e., the output of the whole siamese neural network and the output of the penultimate layer. The output of the whole siamese neural network represents the value of similarity, denoted by $\sigma(\bm{H}_{k}, \bm{H}_{k'}; \boldsymbol{\theta})$, and the output of the penultimate layer represents the \emph{latent factors} of similarity, denoted by $\hat{\sigma}(\bm{H}_{k}, \bm{H}_{k'}; \hat{\boldsymbol{\theta}})$.

Training siamese neural networks should be based on labeled data. Define~$\varrho(\bm{H}_{k}, \bm{H}_{k'}) \in \left\{0, 1\right\}$ as a \emph{label} of the features of network status in two time intervals~$k$ and~$k'$, given by:
	\begin{equation}\label{eq23}
	\varrho(\bm{H}_{k}, \bm{H}_{k'}) = 
	\left\{
             \begin{array}{lr}
            	1, \,\, \text{if} \,\, \Delta_{k} \left( \bm{r}_{k} \right) - \Delta_{k} (\bm{r}_{k'}) > \lambda O^\text{v}; &  \\
             	0, \,\, \text{otherwise}.&  
             \end{array}
	\right.
	\end{equation}
In Eq.~\eqref{eq23}, $\Delta_{k} ( \bm{r}_{k} )$ denotes the network resource usage in time interval~$k$ if the resource reservation is reconfigured, i.e., $a_{k}=0$. $\Delta_{k} ( \bm{r}_{k'} )$ denotes the network resource usage in time interval~$k$ if the resource reservation is not reconfigured, i.e., $a_{k}=1$, and the resource reservation decision from time interval~$k'$ is used, i.e.,~$\bm{r}_{k'}$. If $\Delta_{k} (\bm{r}_{k} ) - \Delta_{k} (\bm{r}_{k'}) > \lambda O^\text{v}$, the network status in two time interval~$k$ and~$k'$ are considered to be ``similar''; Otherwise, the network status in the two time intervals are considered to be ``not similar''. The features of the network status in any two time intervals and the corresponding value of $\varrho(\bm{H}_{k}, \bm{H}_{k'})$ are referred to as a labeled data entry. The goal of training the siamese neural networks is to let $\sigma(\bm{H}_{k}, \bm{H}_{k'}; \boldsymbol{\theta})$ approximate label~$\varrho(\bm{H}_{k}, \bm{H}_{k'})$ by using extensive labeled data entries. The parameters of the siamese neural networks, i.e., $\boldsymbol{\theta}$, are obtained by minimizing the following loss function via gradient descent~\cite{koch2015siamese}:
	\begin{equation}\label{eq24}
			\begin{split}
			\boldsymbol{\theta}^{*} = \,\,&  \arg \min_{ \left\{\boldsymbol{\theta} \right\} } \varrho(\bm{H}_{k}, \bm{H}_{k'}) \log \left(\sigma(\bm{H}_{k}, \bm{H}_{k'}; \boldsymbol{\theta}) \right) + \\
			\,\,& \left(1 - \varrho(\bm{H}_{k}, \bm{H}_{k'}) \right) \log \left(1 - \sigma(\bm{H}_{k}, \bm{H}_{k'};  \boldsymbol{\theta}) \right).
			\end{split}
	\end{equation}

The approximated value of similarity, i.e., $\sigma(\bm{H}_{k}, \bm{H}_{k'}; \boldsymbol{\theta})$, can indicate whether the features of network status in two time intervals are similar or not. However, its information on how much a difference between network status is insufficient for reconfiguring resource reservation.\footnote{The output layer in the siamese neural networks is used for training the siamese neural networks.} Therefore, we use the latent factors of similarity, i.e., $\hat{\sigma}(\bm{H}_{k}, \bm{H}_{k'}; \hat{\boldsymbol{\theta}})$, to determine the value of $a_{k}$. 

\subsection{Closed-loop Resource Reservation Reconfiguration}

\begin{algorithm}[t]
	\caption{Meta$\text{R}^{3}$ Approach}\label{alg3}
	\LinesNumbered
	\textbf{Input:} $\rho$\\
	\textbf{Initialization:} $\boldsymbol{\theta}$, $\hat{\boldsymbol{\theta}}$, $\boldsymbol{\vartheta}$, $k^{\star}$, $\bm{H}_{1}$, $\hat{\sigma}_{1}$\\
	\For{$k = 1,\cdots, K$}
	{		$\sigma_{k}$ $\leftarrow$ Obtain $\hat{\sigma}(\bm{H}_{k}, \bm{H}_{k^{\star}}; \hat{\boldsymbol{\theta}})$; \\
     		$a_{k}$ $\leftarrow$ Determine by Eq.~\eqref{eq27}; \\
     		$\bm{r}_{k}$ $\leftarrow$ Determine by Eq.~\eqref{eq13}; \\
     		$\Delta_{k}(\bm{r}_{k})$, $\hat{\sigma}_{k+1}$ $\leftarrow$ Implement $\bm{r}_{k}$ for time interval~$k$;\\
     		$\boldsymbol{\theta}$, $\hat{\boldsymbol{\theta}}$, $\boldsymbol{\vartheta}$ $\leftarrow$ Train parameters via Eqs.~\eqref{eq24} and~\eqref{eq28};\\ 
     		$k^{\star}$ $\leftarrow$ Update by Eq.~\eqref{eq25};\\

	 }
 \textbf{Output:} $\bm{a}$
\end{algorithm} 

We leverage deep Q learning with deep neural networks, named Q networks, to determine the value of $a_{k}$ given the latent factors of similarity. The state and action in time interval~$k$ are $\hat{\sigma}(\bm{H}_{k}, \bm{H}_{k^{\star}}; \hat{\boldsymbol{\theta}})$ and~$a_{k}$, respectively. For simplicity, let~$\hat{\sigma}_{k}$ denote $\hat{\sigma}(\bm{H}_{k}, \bm{H}_{k^{\star}}; \hat{\boldsymbol{\theta}})$ in the rest of this section. Define a Q-value function to represent the discounted long-term resource usage and cost of making decision~$a_{k}$ in state~$\hat{\sigma}_{k}$, given by:
	\begin{equation}\label{}
		Q(\hat{\sigma}_{k}, a_{k}) =  \sum_{k = 1}^{K}{\rho^{k} \frac{ \Delta_{k} + \lambda o^\text{v}_{k} }{ \sum_{b \in \mathcal B}{\sum_{n \in \mathcal N}{\tilde{x}_{b, k}^{n} }} } } . 
	\end{equation}
where $\rho \in (0, 1)$ is the discount factor. In state~$\hat{\sigma}_{k}$, $a_{k}$ can be determined based on the Q-values as follows:
	\begin{equation}\label{eq27}
		a_{k} = \arg \max_{a \in \left\{0,1\right\}} Q(\hat{\sigma}_{k}, a ),\,\, \forall k \in \mathcal{K}.
	\end{equation}
The Q network with parameter~$\boldsymbol{\vartheta}$ is used to approximate the Q-value function for learning the policy of resource reservation reconfiguration. The parameters~$\boldsymbol{\vartheta}$ of the Q networks are obtained by minimizing the following loss function via gradient descent~\cite{zhou2020deep}:
\begin{equation}\label{eq28}
	\begin{split}
		 \boldsymbol{\vartheta}^{*} = &\arg \min_{\left\{ \boldsymbol{\vartheta} \right\} } \big| \frac{ \Delta_{k} + \lambda o^\text{v}_{k}  }{ \sum_{b \in \mathcal B}{\sum_{n \in \mathcal N}{\tilde{x}_{b, k}^{n} }} } + \rho \max_{a} Q(\hat{\sigma}_{k+1}, a; \boldsymbol{\vartheta})\\
		&   - Q(\hat{\sigma}_{k}, a_{k}; \boldsymbol{\vartheta}) \big|^2.
	\end{split}
\end{equation} 

We summarize the workflow of the Meta$\text{R}^{3}$ approach in Algorithm~\ref{alg3}. Line~4 to Line~5 determine $a_{k}$ at the beginning of time interval~$k$ based on the predicted network status obtained from UDTs. Given $a_{k}$, Line~6 to Line~7 determine and implement the resource reservation decision. At the end of the time interval, data regarding network performance, actual spatial task distributions, and chunk request ratio profiles is collected and stored in UDTs. Given the stored historical data, the siamese neural networks and Q networks can be trained to adapt to dynamic spatial task distribution and chunk request ratio profile in a closed-loop manner.

\section{Performance Evaluation}

\subsection{Simulation Settings}

\begin{table}[t]
\footnotesize 
\centering
\captionsetup{justification=centering,singlelinecheck=false}
\caption{Simulation Parameters}\label{table2}
\begin{tabular}{c|c|c|c}
\hline\hline
 Parameter & Value & Parameter & Value\\
 \hline\hline
 $G^\text{bs}$, $G^\text{nap}$ & 0.9, 2\,Gigacycles/s &$C^\text{bs}$, $C^\text{nap}$& 0.75, 1.5\,GB \\
 \hline
 $\xi_\text{bs}^\text{nap}$, $\xi_\text{nap}^\text{cn}$ & $2.5*10^{-9},$ $3.5*10^{-9}$ & $\xi_\text{bs}^\text{cn}$ & $6*10^{-9}$\\
 \hline
 $\varepsilon_\text{bs}^\text{c}$, $\varepsilon_\text{nap}^\text{c}$,$\varepsilon_\text{cn}^\text{c}$ & 1, 1, 1  & $I$ & 20 \\
 \hline
 $\varepsilon_\text{bs}^\text{s}$, $\varepsilon_\text{nap}^\text{s}$,$\varepsilon_\text{cn}^\text{s}$  & 0.5, 0.8, 1 & $L$ & 0.15\,GB \\
 \hline
 $\eta^\text{nap}$,  $\eta^\text{cn}$ & $5*10^{-9},$ $9*10^{-9}$ & $\tau^\text{p}$ & 0.5\,s \\
 \hline
 $w^\text{s}$, $w^\text{c}$, $w^\text{o}$ & $0.5*10^{-7}$, 1, 1& $\lambda$ & 12 \\
 \hline
\end{tabular}
\end{table}

The simulated multi-tier network consists of one S-CN, two S-NAPs, and four to ten S-BSs. There are 600 UTs with different trajectories. Based on the average time within the coverage of each BS for each UT, these UTs are grouped into two to four groups. The input data size, computing workload, and the size of computing results of each computing task are set to 2\,MB, 4\,Megacycles/s, and 15\,MB, respectively~\cite{cheng2019space}. The network resource usage and the resource capacity of servers at the same tier can be different. The average network resource usage, average resource capacity of servers, and other parameters are listed in Table~II. 

For the siamese neural networks illustrated in Fig.~\ref{nn}, we use 3 fully connected layers with (64, 64, 32) neurons as a embedding layer. The features of network status $\bm{H}_{k}$ and $\bm{H}_{k'}$ are fed to two embedding layers separately, each with the same structure. The merging layer merges the outputs of the two embedding layers based on Euclidean distance, followed by the penultimate layer with 16 neurons and the output layer with 1 neuron. For the Q networks, we adopt 4 fully connected layers with 128, 512, 128, 32 neurons, respectively. We adopt the RMSprop optimizer and the Adam optimizer for training the siamese neural networks and the Q networks, respectively.

\subsection{Performance of Group-based Resource Reservation}

\begin{figure}[t]
	\centering
  	\includegraphics[width = 0.40\textwidth]{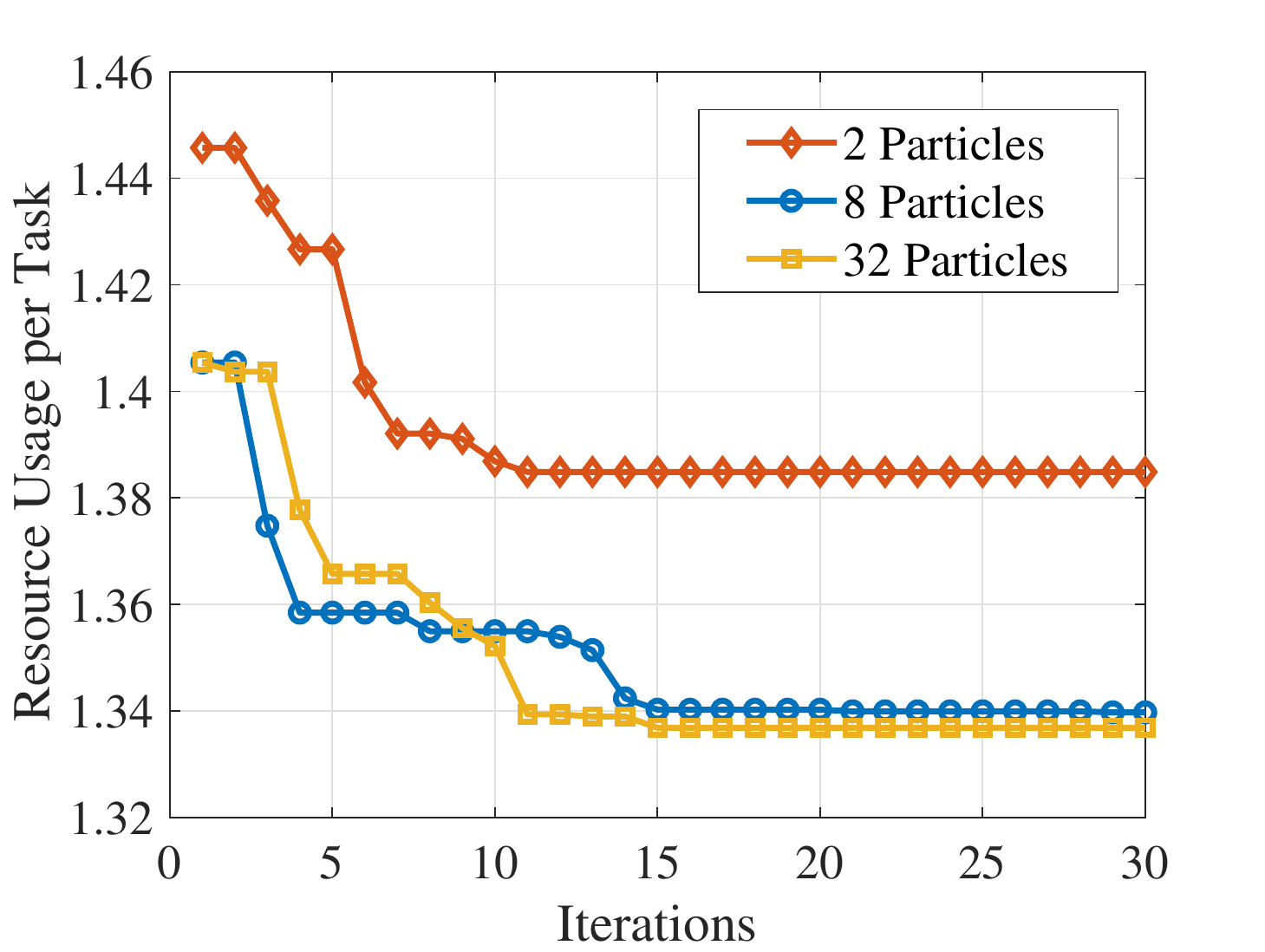}
  	\caption{Convergence performance of the proposed RR algorithm.}\label{fig1}
\end{figure}

The convergence performance of the proposed RR algorithm for group-based resource reservation is shown in Fig.~\ref{fig1}. Given the same spatial task distribution and configuration of all servers, we conduct the simulation with 2, 8, and 32 particles for 30 iterations. The proposed algorithm converges after 10, 14, and 16 iterations, respectively. With more particles, the algorithm achieves better performance at the cost of computation complexity.

\begin{figure}
  \centering
    \subfigure[Uneven spatial task distribution.]
    { \includegraphics[width=0.4\textwidth]{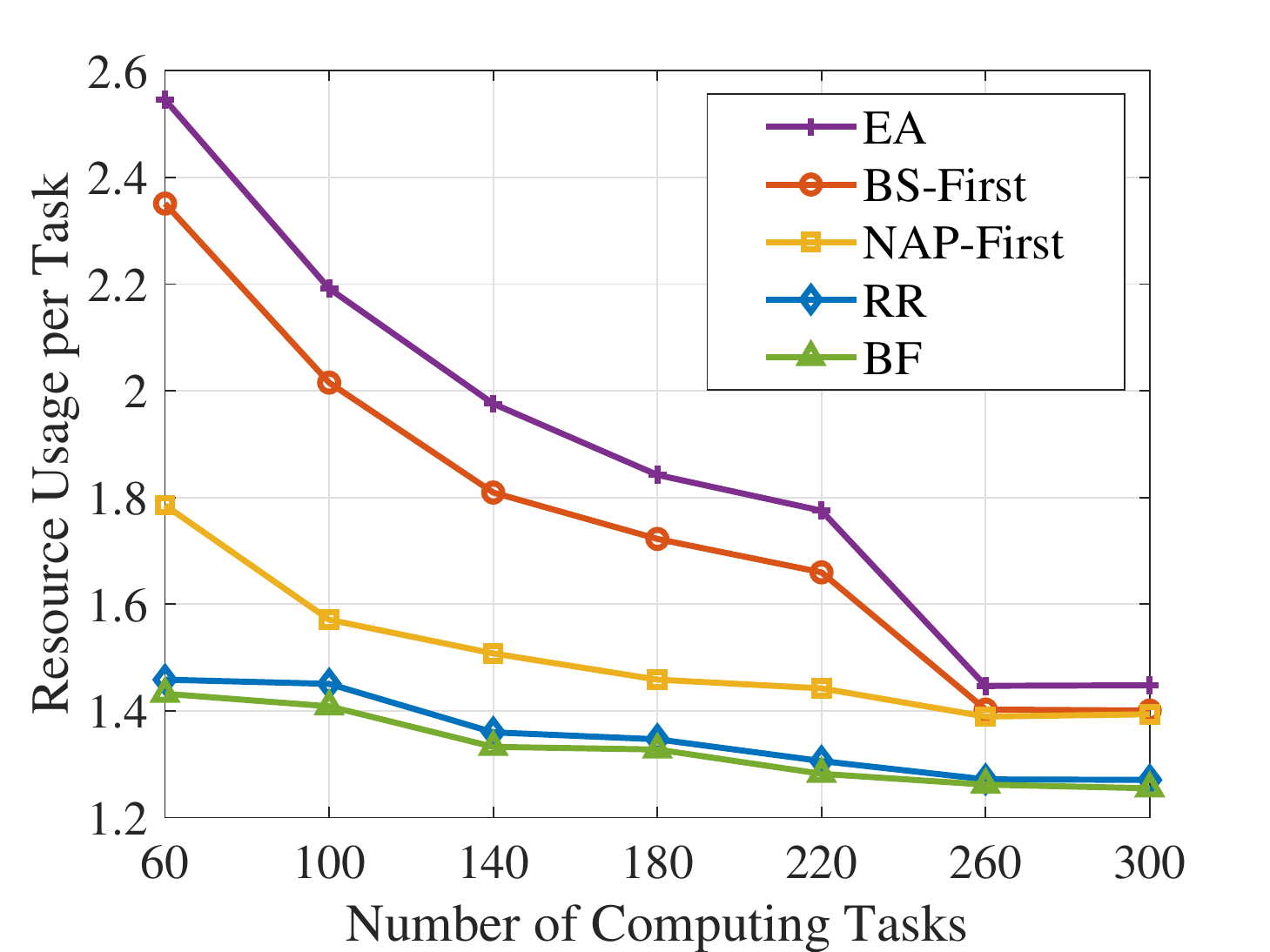}\label{fig21} }
	\subfigure[Even spatial task distribution.]
	{ \includegraphics[width=0.4\textwidth]{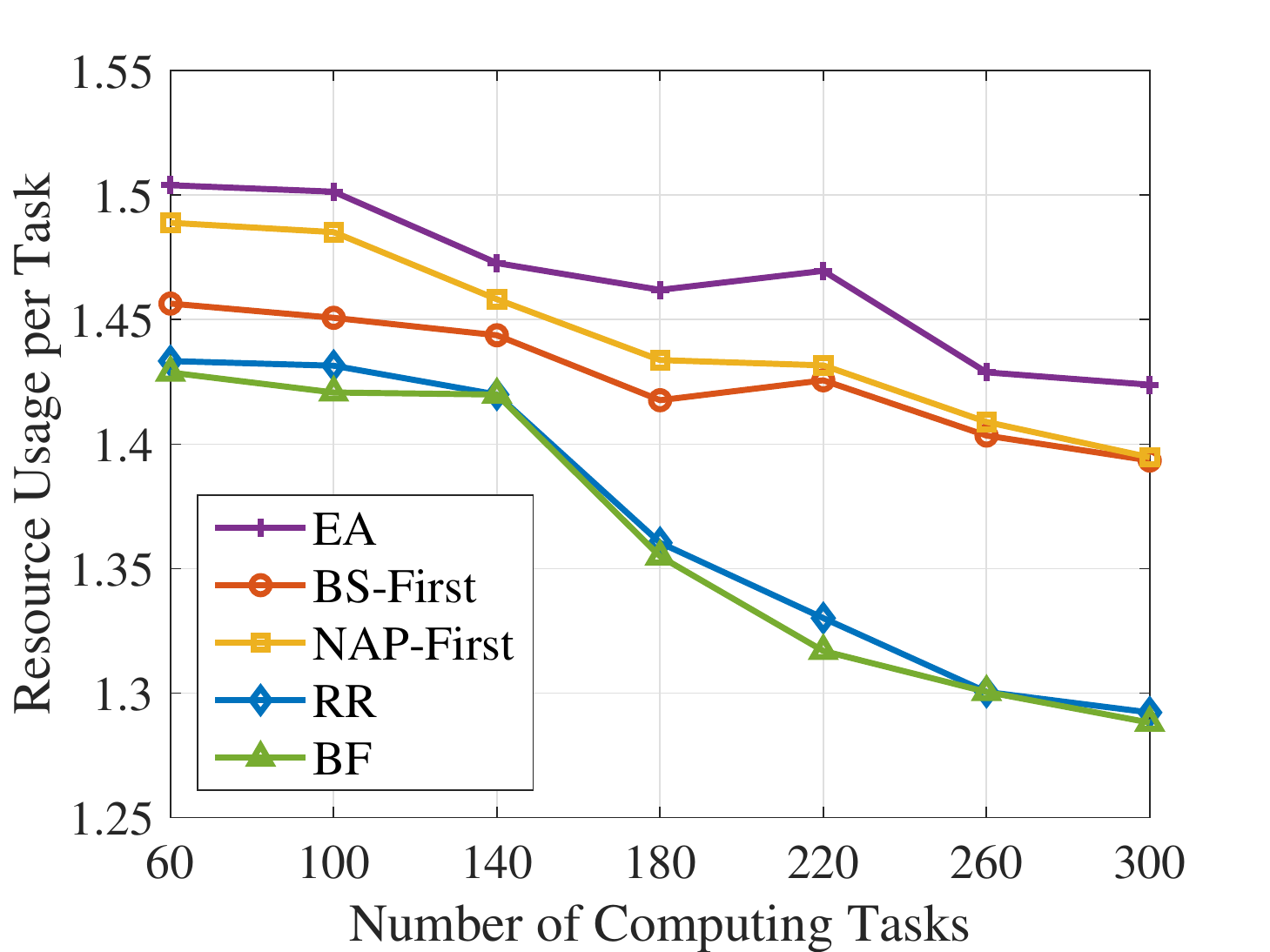}\label{fig22} }
  \caption{Network resource usage per computing task under even and uneven spatial task distributions.}\label{fig2}
\end{figure}

Figures~\ref{fig21} and~\ref{fig22} compare the network resource usage per computing task, versus the load of computing tasks, of the proposed RR algorithm and the benchmark algorithms. We adopt four benchmark algorithms: (1) BS-first, which assigns computing tasks to S-BSs and reserves storage and computing resources on S-BSs first as much as possible, then on S-NAPs, and last on the S-CN. (2) NAP-first, which assigns computing tasks to S-NAPs and reserves storage and computing resources on S-NAPs first as much as possible, then on S-BSs, and last on the S-CN. (3) EA, which assigns computing tasks and reserves storage and computing resources on S-BS, S-NAP, and S-CN with equal priority. (4) BF, which is a brute force algorithm to find the global optimum. We have the three following observations. First, for both even and uneven task distribution, the network resource usage per computing task of the RR algorithm is close to the global optimum and much lower compared to the benchmark algorithms (except the BF algorithm). Second, the network resource usage per computing task decreases as the load of computing tasks generated in the network increases. This is because the storage resource usage per task decreases as the load of computing tasks requesting the same stored chunk increases. Third, the performance gap between the RR algorithm and the benchmark algorithms (except the BF algorithm) under uneven spatial task distribution is larger than under even spatial task distribution. With uneven spatial task distribution, the optimal resource reservation may be different for servers at the same tier. The RR algorithm can differentiate resource reservation decisions for different servers at the same tier using group-based spatial task distribution.

\subsection{Performance of Closed-loop Resource Reservation Reconfiguration}

\begin{figure*}
  \centering
    \subfigure[Convergence performance of Meta$\text{R}^{3}$.]
    { \includegraphics[width=0.31\textwidth]{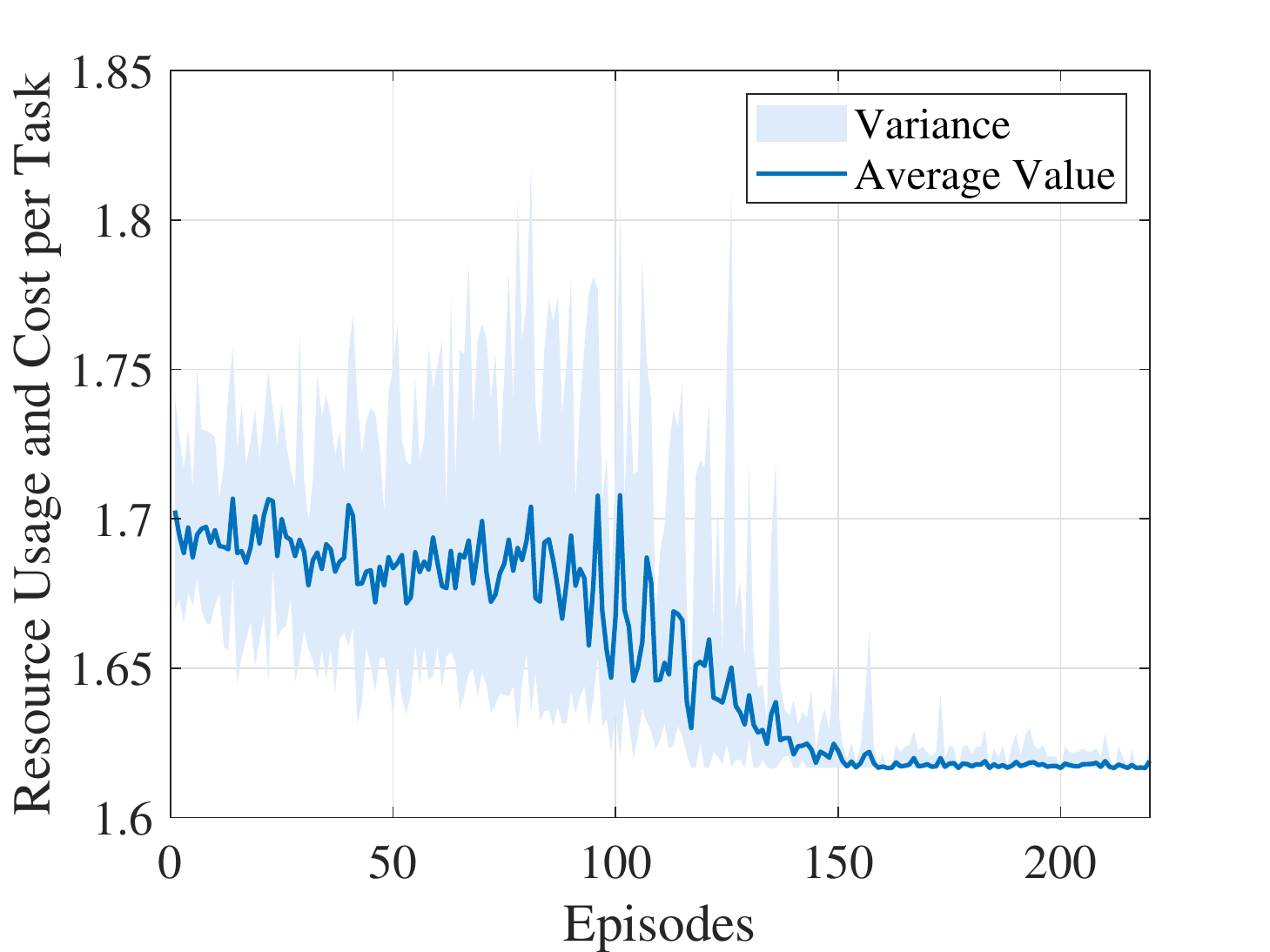}\label{fig4} }
	\subfigure[Performance Comparison between Meta$\text{R}^{3}$ and DQN.]
	{ \includegraphics[width=0.31\textwidth]{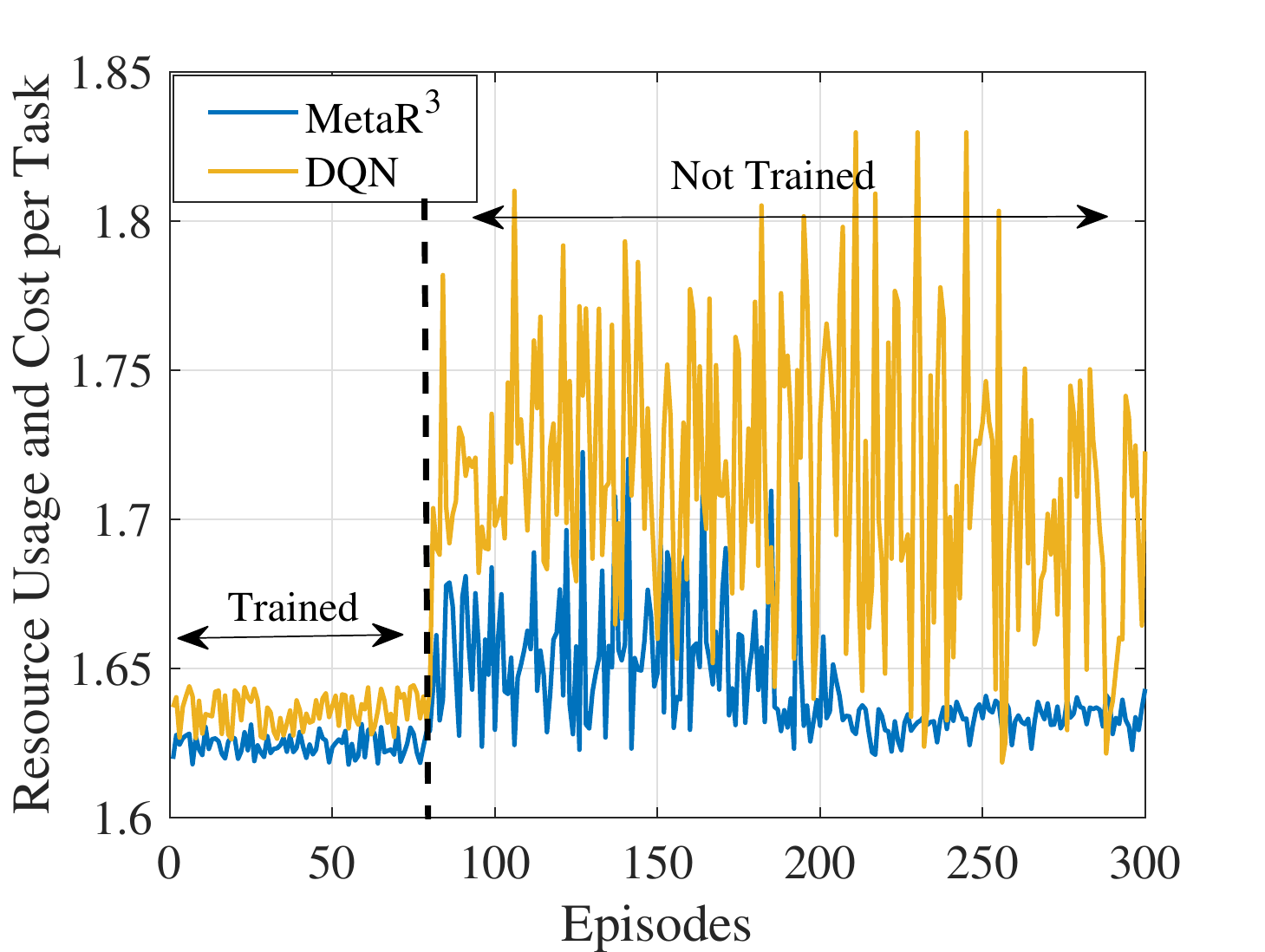}\label{fig5} }
		\subfigure[The impact of network dynamics.]
	{ \includegraphics[width=0.31\textwidth]{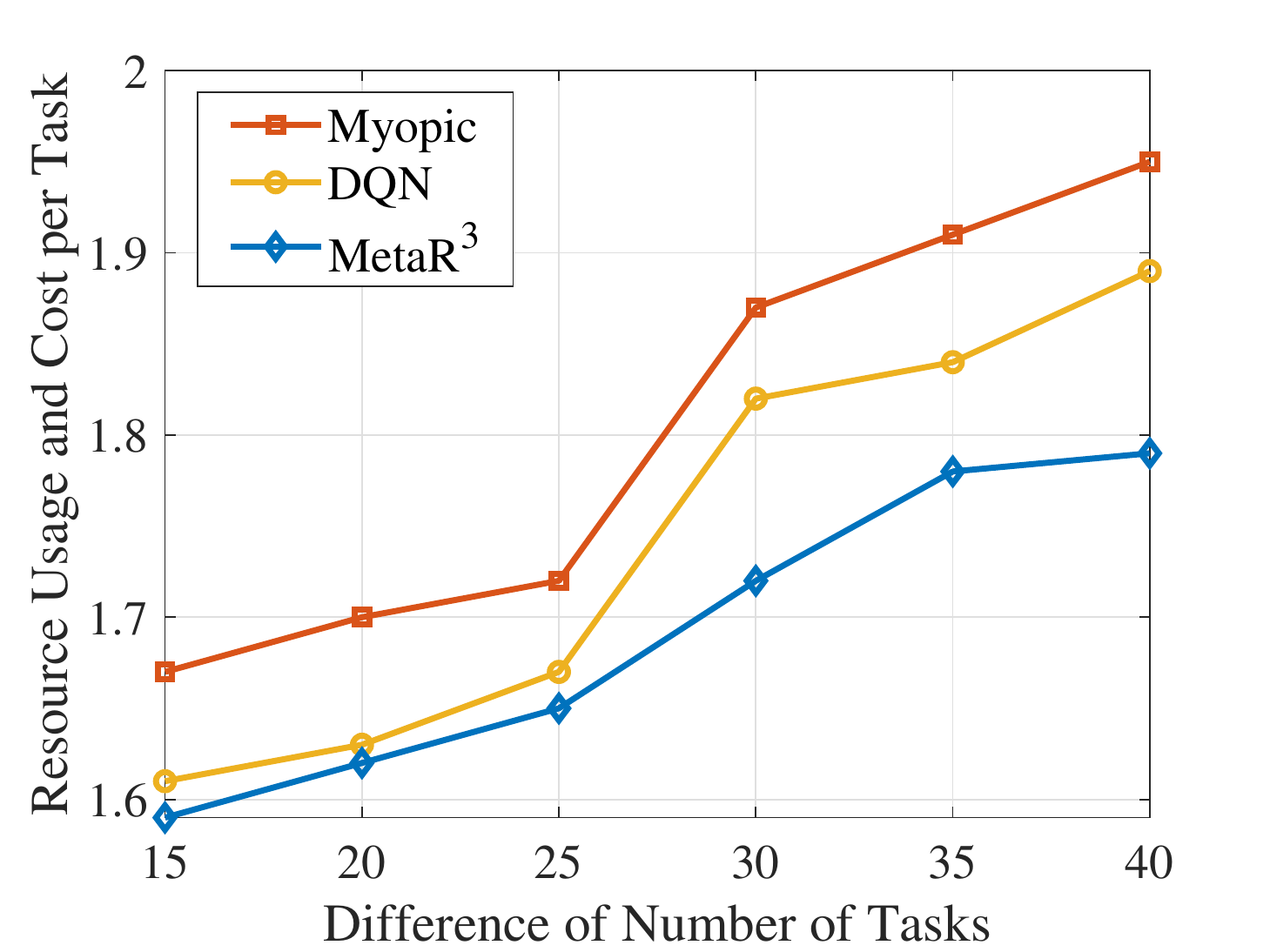}\label{fig6} }
  \caption{Performance of Meta$\text{R}^{3}$ in the weighted sum of network resource usage and cost from reconfiguring resource reservation per computing task.}\label{fig456}
\end{figure*}

A training dataset that includes spatial task distributions in 80 time intervals is created. Conducting the simulation for all 80 spatial task distributions in the training dataset is referred to as one episode. We conduct 10 simulations on the dataset, and each simulation includes 220 episodes. In Fig.~\ref{fig4}, the smooth solid line is the average result over 10 simulations, while the spikes in the background represent the corresponding variance. Fig.~\ref{fig4} shows that the proposed Meta$\text{R}^{3}$ algorithm can converge and find a policy of resource reservation reconfiguration given a fixed network environment. 

In Fig.~\ref{fig5}, we compare the convergence performance of the Meta$\text{R}^{3}$ approach with that of a deep Q Learning (DQN) based algorithm, labeled as ``DQN''. In DQN, the group-based spatial task distribution is used as the state to determine the value of $a_{k}$. We create two datasets with different spatial task distributions. One training dataset is used to train the neural networks in advance, and one evaluation dataset is used to evaluate the convergence performance of Meta$\text{R}^{3}$. The evaluation dataset reveals network status from unknown network environments. Note that Meta$\text{R}^{3}$ keeps training the siamese neural networks and the Q networks in unknown network environment due to the closed-loop reconfiguration of resource reservation. We observe that Meta$\text{R}^{3}$ achieves lower network resource usage and lower cost from reconfiguring resource reservation per computing task, and also converges in fewer episodes in unknown network environment compared with the DQN algorithm. This is because Meta$\text{R}^{3}$ captures the similarity of network status, instead of learning the variation of network status, to determine~$a_{k}$ even though the current network status is unknown.

Figure~\ref{fig6} shows the performance in the weighted sum of the network resource usage and the cost from reconfiguring resource reservation per computing task versus the average difference in the load of computing tasks in adjacent two time intervals. When the average difference in the load of computing tasks in adjacent two time intervals increases, spatial task distribution changes faster. The benchmark algorithm, labeled as ``Myopic'', determines whether to reconfigure resource reservation or not in each time interval without considering the long-term impact. We observe that the performance gaps between Meta$\text{R}^{3}$ and ``DQN'' and ``Myopic'' algorithms increase with the average difference in the load of computing tasks in adjacent two time intervals since the similarity capture features of Meta$\text{R}^{3}$ can reduce the state space for finding a good policy of resource reservation reconfiguration in dynamic network environments, which improves learning efficiency. 

	\begin{figure}[t]
		\centering
	  	\includegraphics[width=0.4\textwidth]{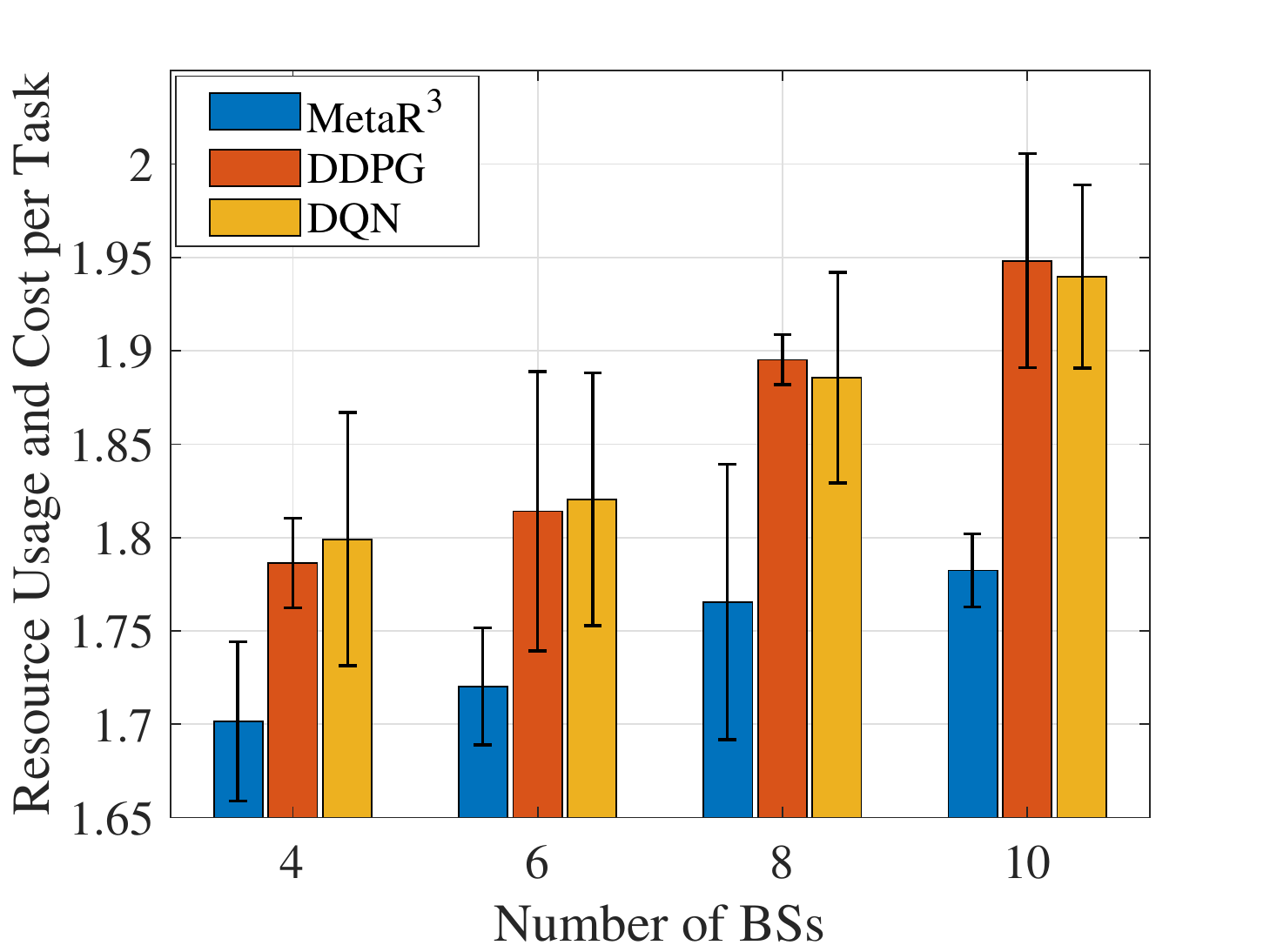}
	  	\caption{Resource usage and cost per computing task of the Meta$\text{R}^{3}$, DDPG, and DQN-based algorithms.}\label{figR1}
	\end{figure}
In Fig.~\ref{figR1}, we compare the performance of the proposed Meta$\text{R}^{3}$ algorithm with that of two popular RL algorithms, i.e., deep deterministic policy gradient (DDPG)-based (labeled as ``DDPG'' ) and DQN-based algorithms in different network environments~\cite{zhang2021dynamic,nouruzi2022online}. Specifically, we conduct simulations of the three algorithms with different numbers of BSs and average the resource usage and cost per computing task over three independent simulations. We can observe that the proposed Meta$\text{R}^{3}$ algorithm outperforms the DDPG- and DQN-based algorithms. This is because both DDPG- and DQN-based algorithms use network status as states. When the network status has a large dimensionality and network environments are highly dynamic, finding the optimal resource reservation reconfiguration policy is challenging for the DDPG and DQN-based algorithms. In contrast, the proposed Meta$\text{R}^{3}$ algorithm can capture the similarity of network status in consecutive time intervals. Since the similarity is low-dimensional, using similarities as states has advantages on finding a proper policy of resource reservation reconfiguration, particularly in complicated network environments. Therefore, the proposed Meta$\text{R}^{3}$ algorithm achieves better network performance than DDPG- and DQN-based algorithms.

\section{Conclusion and Future Work}

In this paper, we have designed DT-empowered network planning for supporting stateful applications in multi-tier computing and proposed two approaches to enable group-based multi-resource reservation and closed-loop resource reservation reconfiguration. Our study focuses on minimizing the long-term network resource usage and the cost from reconfiguring resource reservation. The results have demonstrated that DT-empowered network planning can support UTs with diverse characteristics and adapt to dynamic network environments. In addition, the Meta-learning-based approach can exploit data contained in DTs to facilitate closed-loop network planning. Overall, we have demonstrated the essential role that DTs can play in network planning for 6G. In the future, we will improve the flexibility of DTs by differentiating and optimizing DTs for various applications or UTs.

\appendix

\subsection{Effective Request Ratio for S-NAPs}\label{app1} 

For S-BS~$e$ located at BS~$b$, S-BS~$e$ sorts $|\mathcal{I}_{e, k}|$ chunks with largest values of $p_{b, k}^{i}$ in time interval~$k$. Let $J_{e,k}^{i}$ denote the order of chunk~$i$ among the chunks with largest values of $p_{b, k}^{i}$ in set $\mathcal{I}_{e,k}$, and denote by $\mathcal{I}^{(J_{e,k}^{i})}_{e,k} \subseteq \mathcal{I}_{e,k}$ the set of $J_{e,k}^{i}$ chunks with largest values of~$p_{b,k}^{i}$. We assume that the computing tasks requesting any chunk in~$\mathcal{I}_{e,k}$ are assigned to the S-BS as much as possible while not violating the communication and computing resource capacities. Given different values of~$g_{e,k}$ for S-BS~$e$, the load of computing tasks assigned to S-BS~$e$ may be different. For S-BS~$e$ co-located with BS~$b$, the overall load of computing tasks requiring any chunk in set~$\mathcal{I}^{(J_{e,k}^{i})}_{e,k}$ is given by $\sum_{i \in \mathcal{I}^{(J_{e,k}^{i})}_{e,k}}{ \tilde{x}_{b,k}^{(i)} }$ where $\tilde{x}_{b,k}^{(i)} = p_{b,k}^{i} \sum_{n \in \mathcal{N}}{x_{b,k}^{n}}$ is the load of computing tasks requiring chunk~$i$ in the coverage of BS~$b$ in time interval~$k$. The load of computing tasks that request chunk~$i \in \mathcal{I}_{e,k}$ in the coverage of BS~$b$ and are not assigned to S-BS~$e$ in time interval~$k$, denoted by $P_{b,e,k}^{i}$, is as follows:
	\begin{equation}\label{eq30}
		\begin{aligned}
			& P_{b,e,k}^{i} = \\
			& \left\{ \begin{array}{l}
			 	0,\,\, \text{if}\,\, \sum_{i \in \mathcal{I}^{(J_{e,k}^{i})}_{e,k}}{ \tilde{x}_{b,k}^{(i)} } \leq M_{e};\\ 
				\min \left\{\tilde{x}_{b,k}^{(i)}, \sum_{i \in \mathcal{I}^{(J_{i,e,k})}_{e,k}}{ \tilde{x}_{b,k}^{(i)} } - M_{e} \right\},\,\, \text{otherwise}.
			\end{array} \right.
		\end{aligned}
	\end{equation}
where $M_{e} =  \lfloor \frac{\tau^\text{p} C_{e}}{\alpha \beta}  \rfloor $ is the maximum load of computing tasks that can be assigned to S-BS~$e$ with satisfying the computing capacity. In Eq.~\eqref{eq30}, if S-BS~$e$ has sufficient computing resource for executing all computing tasks requiring chunk~$i$, i.e., $\sum_{\mathcal{I}^{(J_{e,k}^{i})}_{e,k}}{ \tilde{x}_{b,k}^{(i)} } \leq M_{e}$, no computing task requiring chunk~$i$ needs to be assigned to an S-NAP or the S-CN; Otherwise, a certain load of computing tasks requiring chunk~$i$ cannot be processed at S-BS~$e$, which should be assigned to an S-NAP or the S-CN.

The overall load of computing tasks that request chunk~$i$ and are not assigned to any S-BS~$e$ within the service coverage of S-NAP~$e'$ in time interval~$k$ is given by $\sum_{b \in \mathcal{B}_{e'}}{P_{b,e,k}^{i} }$. The effective request ratio of chunk~$i$ for S-NAP~$e'$ in time interval~$k$ is as follows: 
	\begin{equation}\label{}
		q^{i}_{e',k} = \frac{  \sum_{b \in \mathcal{B}_{e'}}{P_{b,e,k}^{i} }  }{  \sum_{b \in \mathcal{B}_{e'}}{\sum_{\mathcal{N}}{ \tilde{x}_{b,k}^{n} }}  }, \,\, \forall i \in \mathcal{I}, e' \in \mathcal{E}^\text{nap}, k \in \mathcal{K}.
	\end{equation}
S-NAP~$e'$ stores $|\mathcal{I}_{e', k}|$ chunks with the largest values of $q^{i}_{e',k}$.

\bibliographystyle{IEEEtran}
\bibliography{./ref}

\end{document}